\documentclass[aps,pra,twocolumn,groupedaddress,longbibliography]{revtex4-1}
\usepackage{amsmath,bm}	
\usepackage[dvipdfmx]{graphicx} 
\usepackage{color}
\begin{document}
\title{Hybrid method of plane-wave and cylindrical-wave expansions for distributed Bragg-reflector pillars: formalism and its application to topological photonics }
\author{Tetsuyuki Ochiai}
\affiliation{Research Center for Functional Materials, National Institute for Materials Science (NIMS), Tsukuba 305-0044, Japan}
\begin{abstract} 
A hybrid computational method of plane-wave  and cylindrical-wave expansions for distributed Bragg-reflector (DBR) pillars is proposed. The plane-wave expansion is employed to represent the one-dimensional periodic structure of the DBR. The cylindrical-wave expansion is employed to describe the scattering by circular pillars with the DBR structure inside. This formalism enables us to calculate the radiation fields, $t$-matrices, scattering cross sections, photonic band structures, and quality factors of the DBR pillars. Furthermore, optical properties of arrayed DBR pillars are also investigated with the aid of the multiple-scattering method.   Using this formalism, we demonstrate explicitly that high $Q$ photonic band modes including the so-called bound states in continuum are obtained both in isolated and arrayed DBR pillars.    
We also present a novel formation of gapless Dirac-cone surface states in a three-dimensional photonic crystal composed of a two-dimensional periodic arrangement of core-shell DBR pillars.   
\end{abstract}
\date{\today}
\maketitle
\section{introduction}

Channel waveguides with one-dimensional (1D) periodic modulations in refractive index, such as two-dimensional (2D) photonic crystal (PhC) slab defect waveguides \cite{PhysRevB.62.8212,PhysRevLett.87.253902}, optical fiber gratings \cite{erdogan1997fiber}, and distributed Bragg reflector (DBR) pillars \cite{santori2002indistinguishable} are very important as a platform of slow light \cite{baba2008slow}, waveguide quantum electrodynamics \cite{PhysRevLett.98.153003}, chiral quantum optics \cite{PhysRevLett.121.053901}, and  multi-harmonic generation \cite{corcoran2009green}.  
They are characterized by well-defined dispersion relations of light outside the light cone, having band gaps at the boundary of the Brillouin zone. The band gaps act as stop bands, in which light cannot propagate in the waveguides. Inside the light cone, quasi-guided modes are embedded in the radiation continuum, resulting in the Fano resonance \cite{PhysRev.124.1866} in light transport spectra. Such features are typical consequences of strongly modulated optical density of states (DOS) of the systems, giving rise to a variety of nontrivial phenomena relevant to light-matter interaction.

Here, we focus on circular DBR pillars (as well as coaxial optical fiber gratings of circular shapes), and propose an optimal computation scheme for them utilizing the circular symmetry.  The reason why we choose these structures is that they have the  simplest geometries, having periodic dielectric functions that depend solely on $z$ coordinate (taken to be parallel to the waveguides) inside the structures. This feature together with the circular symmetry enables us to employ a combination of the plane-wave and cylindrical-wave expansions. If the periodic modulation in the $z$ direction is absent, the analytic solution via the cylindrical wave expansion is available as the Mie scattering \cite{hulst1957light}. The plane-wave expansion is a first choice to describe the periodic modulation. Moreover, since the DBR structure with infinite extent in plane has the analytic solutions, we can go beyond the plane-wave expansion with inevitable Gibbs phenomena, via the eigenmode expansion \cite{botten1981finitely}. We should note that a similar computational method but without the eigenmode expansion  was proposed by Li and Engheta \cite{PhysRevB.74.115125}.

Conventional methods to study such DBR pillars are general-purpose ones such as the finite-difference time-domain method \cite{Taflove-FDTD-book} and finite-element method \cite{jin2015finite}. Although these methods are versatile, more optimized methods such as the one proposed in this paper are definitely in order. This is because optimized methods 1) generally have higher accuracy, 2) provide cross checks to the conventional methods, and 3) can be adapted to a wider class of materials having circular-symmetric permittivity tensor (uniaxial or magneto-optical ones) without reducing the numerical accuracy.

The proposed hybrid method allows us to investigate fundamental optical properties of isolated DBR pillars.  We demonstrate its potential by evaluating the photonic band structure of a DBR pillar, inside and outside the light cone with high accuracy.  In particular, bound states in continuum \cite{hsu2016bound} due to a symmetry mismatch are clearly identified. 
Moreover, the multiple-scattering formalism \cite{twersky1952multiple} can be merged to the hybrid method. We also present a formalism to deal with 1D and 2D periodic arrays of DBR pillars, and demonstrate their band structures.

As a nontrivial application of the method, we consider domain walls in a three-dimensional (3D) PhC composed of a 2D periodic arrangement of DBR pillars. Such a 3D structure can exhibit a novel surface-state formation with Dirac-cone dispersion as shown in this paper.  The Dirac-cone surface states are inherent in certain topological systems, particularly, topological insulators \cite{Hasan2010}. Therefore, our method can deal with topological photonics \cite{lu2014topological,ozawa2018topological} in a unique manner.

This paper is organized as follows. 
In Sec. II, we present a theory of light scattering by isolated DBR pillars. Section III is devoted to describe the multiple scattering of light in parallel arrays of DBR pillars. In Sec. IV, we present the gapless surface-states formation in 3D PhCs composed of core-shell DBR pillars. Finally in Sec. V, we summarize the results.

\section{Isolated DBR pillars}
\begin{figure}
\begin{center}
\includegraphics[width=0.4\textwidth]{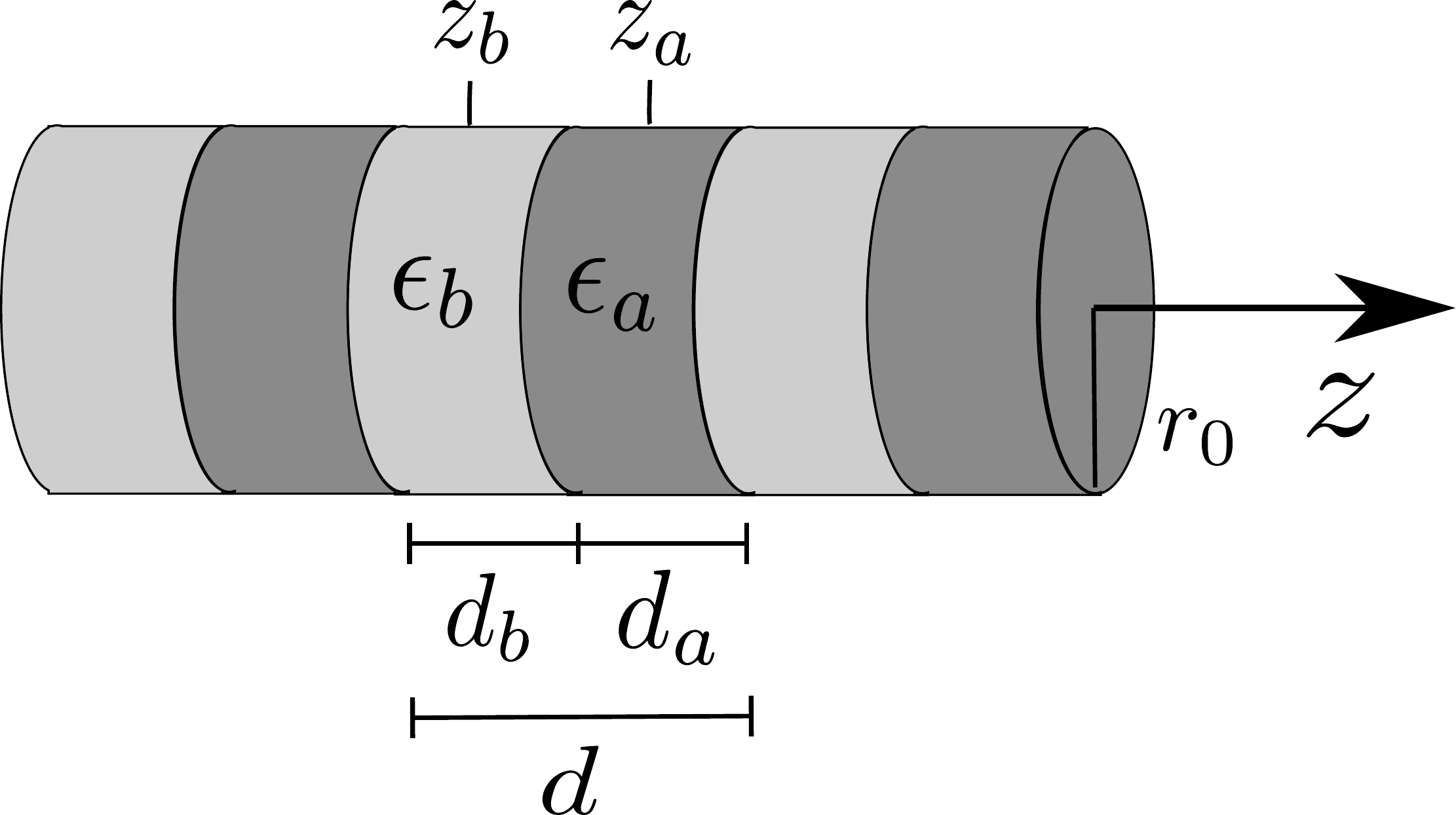}
\end{center}
\caption{\label{Fig_fibgra} Schematic illustration of the system under study. It consists of a periodic arrangement of two kinds of finite-height pillars.  One has permittivity $\epsilon_a$ and height $d_a$. The other has $\epsilon_b$ and $d_b$, respectively. Both pillars have the same radius $r_0$. As a whole, the system has the one-dimensional periodicity with period $d=d_a+d_b$.   
} 
\end{figure}

Let us consider a light-scattering problem by a DBR pillar composed of alternating finite-height pillars with the circular cross section.   The period is denoted as $d$, and the pillar axis is taken to be the $z$ direction. A schematic illustration of the system under study is shown in Fig. \ref{Fig_fibgra}. 
Since there is the translational invariance in the $z$ direction, we introduce Bloch momentum $k_z$.

Suppose that an incident plane-wave light of angular frequency $\omega$, momentum $k_z^0$, azimuthal angle $\phi_0$ of wave vector, and polarization ${\bm p}_0$ is coming from the outside.  
Its electric field is given by 
\begin{align}
&{\bm E}^0({\bm x})={\bm p}_0 {\rm e}^{{\rm i}{\bm k}^0\cdot{\bm x}}, \\
&{\bm k}^0 =\left(\begin{array}{c}
\lambda_0\cos\phi_0\\
\lambda_0\sin\phi_0\\
k_z^0
\end{array}\right), \quad 
\lambda_0=\sqrt{q^2-(k_z^0)^2}, \quad q=\frac{\omega}{c}. 
\end{align}
Here, we omit a time-harmonic dependence of the radiation field, namely, 
\begin{align}
{\bm F}({\bm x},t)=\Re[{\bm F}({\bm x}){\rm e}^{-{\rm i}\omega t}] \quad ({\bm F}={\bm E},{\bm B},{\bm D},{\bm H}),   
\end{align}
and the complex field ${\bm F}({\bm x})$ is considered throughout the paper.

The incident light is then Bragg scattered by the DBR pillar, and the resulting scattered light is a superposition of propagating and evanescent waves of momentum $k_z+g_z$ [being $g_z(=2\pi{\bm Z}/d)$ a reciprocal lattice] in the $z$ direction. 
The incident and induced (scattered) electric fields outside the DBR pillar are written as 
\begin{align}
&{\bm E}^{0}({\bm x})=\sum_{g_z,l,l',\beta}J_l(\lambda_{g_z}\rho){\rm e}^{{\rm i}l\phi}{\rm e}^{{\rm i}(k_z+g_z)z}[{\bm P}_{g_z}^{\beta}]_{ll'}\psi_{l'g_z}^{\beta 0},\\
&{\bm E}^{\rm ind}({\bm x})=\sum_{g_z,l,l',\beta}H_l(\lambda_{g_z}\rho){\rm e}^{{\rm i}l\phi}{\rm e}^{{\rm i}(k_z+g_z)z}[{\bm P}_{g_z}^{\beta}]_{ll'}\psi_{l'g_z}^{\beta {\rm ind}},\\
&\psi_{lg_z}^{\beta 0}=\delta_{g_zg_z^0}\sum_{l'}{\bm p}_0\cdot [\tilde{\bm P}_{g_z^0}^\beta]_{ll'}i^{l'}{\rm e}^{-{\rm i}l'\phi_0}\quad (k_z^0=k_z+g_z^0),\\ 
&\lambda_{g_z}=\sqrt{q^2-(k_z+g_z)^2}.   
\end{align}
Here, $J_l$ is the Bessel function of integer order $l$, $H_l$ is the Hankel function of the first kind and integer order $l$, $(\rho,\phi,z)$ is the cylindrical coordinate, and $[{\bm P}_{g_z}^{\beta}]_{ll'}$ ($\beta=M,N$) is the transformation matrix of the vector cylindrical waves \cite{Ohtaka:U:A::57:p2550-2568:1998}:
\begin{align}
&[{\bm P}_{g_z}^{M}]_{ll'}=\left(\begin{array}{l}
\frac{i}{2}(\delta_{l,l'+1}+\delta_{l,l'-1}) \\
\frac{1}{2}(\delta_{l,l'+1}-\delta_{l,l'-1}) \\
0 
\end{array}\right),\\ 
&[{\bm P}_{g_z}^{N}]_{ll'}=\left(\begin{array}{l}
-\frac{i(k_z+g_z)}{2q}(\delta_{l,l'+1}-\delta_{l,l'-1})\\
-\frac{k_z+g_z}{2q}(\delta_{l,l'+1}+\delta_{l,l'-1})\\
\frac{\lambda_{g_z}}{q}\delta_{l,l'} 
\end{array}\right),\\
& \tilde{\bm P}_{g_z}^{M}=-{\bm P}_{g_z}^{M},\quad  \tilde{\bm P}_{g_z}^{N}={\bm P}_{g_z}^{N}. 
\end{align} 
The $M$ and $N$ components stand for two transverse degrees of vector cylindrical waves \cite{Stratton-EMbook}.

The $t$-matrix that describes the light scattering by the DBR pillar is defined by 
\begin{align}
\psi_{lg_z}^{\beta {\rm ind}}=\sum_{g_z'\beta'} [t_l]_{g_zg_z'}^{\beta\beta'}\psi_{lg_z'}^{\beta' 0}. \label{Eq_tmat}
\end{align} 
Here, we have the rotational symmetry about the cylindrical axis, so that the $t$-matrix is diagonal with respect to angular momentum index $l$.   
Such a $t$-matrix is also obtained in optical fiber gratings \cite{erdogan1997fiber} and structured optical fibers with a periodic array of holes inside \cite{ochiai2010imitating}.

In what follows we derive the $t$-matrix in two different schemes: the Fourier modal basis and layer-by-layer basis.

\subsection{Fourier modal basis}

In this scheme, the dielectric function $\epsilon$ and radiation field ${\bm F}$ inside the DBR pillar are expanded in the Fourier series as 
\begin{align}
&\epsilon(z)=\sum_{g_z} {\rm e}^{{\rm i}g_zz}\epsilon_{g_z}, \\
&\frac{1}{\epsilon(z)}=\sum_{g_z} {\rm e}^{{\rm i}g_zz}\eta_{g_z}, \\
&{\bm F}({\bm x})=\sum_{g_z} {\rm e}^{{\rm i}(k_z+g_z)z}{\bm F}_{g_z}({\bm \rho}), 
\end{align}
and the Fourier coefficients ${\bm F}_{g_z}({\bm \rho})$ are to be solved.

Inside the DBR pillar, eigenmodes are classified into two categories. One is the $M$ polarization and the other is the $N$ polarization. In the former polarization, the electric displacement ${\bm D}$ is given by  
\begin{align}
&{\bm D}({\bm x})=-\hat{z}\times{\bm \nabla}\psi^M({\bm x}),\\ 
&\psi^M({\bm x})=\sum_{lg_z} \psi_{lg_z}^M(\rho) {\rm e}^{{\rm i}l\phi}{\rm e}^{{\rm i}(k_z+g_z)z},\\
&\tilde{\psi}_{lg_z}^M(\rho)=\sum_{g_z'}\eta_{g_z-g_z'}\psi_{lg_z'}^M(\rho),\\
&\left(\frac{\partial^2}{\partial \rho^2} + \frac{1}{\rho}\frac{\partial}{\partial \rho}-\frac{l^2}{\rho^2}\right)\tilde{\psi}^M_{lg_z}(\rho) +\sum_{g_z'}M_{g_zg_z'}\tilde{\psi}^M_{lg_z'}(\rho)=0, \label{Eq_Mpol}\\
&M_{g_zg_z'}=q^2\epsilon_{g_z-g_z'}-(k_z+g_z)^2\delta_{g_zg_z'},
\end{align}
and the magnetic field ${\bm B}$ is obtained by the Faraday law $\nabla\times {\bm E}=i\omega {\bm B}$ with ${\bm E}={\bm D}/(\epsilon_0\epsilon(z))$. By diagonalizing the second term in Eq. (\ref{Eq_Mpol}), $\tilde{\psi}_{lg_z}$ is expanded as 
\begin{align}
&\tilde{\psi}_{lg_z}^M(\rho)=\sum_\alpha c_{l\alpha}^M J_l(\sigma_\alpha^M\rho) \tilde{\psi}_{\alpha g_z}^M,\label{Eq_Mpolexp}\\  
&\sum_{g_z'}M_{g_zg_z'}\tilde{\psi}_{\alpha g_z'}^M=(\sigma_\alpha^M)^2 \tilde{\psi}_{\alpha g_z}^M. 
\end{align}

Similarly, in the $N$ polarization, the magnetic field is given by 
\begin{align}
&{\bm B}({\bm x})=-\hat{z}\times{\bm \nabla}\psi^N({\bm x}),\\ 
&\psi^N({\bm x})=\sum_l \psi_{lg_z}^N(\rho) {\rm e}^{{\rm i}l\phi}{\rm e}^{{\rm i}(k_z+g_z)z},\\
&\left(\frac{\partial^2}{\partial \rho^2} + \frac{1}{\rho}\frac{\partial}{\partial \rho}-\frac{l^2}{\rho^2}\right)\psi_{lg_z}^N(\rho)+\sum_{g_z'}N_{g_zg_z'}\psi_{lg_z'}^N(\rho)=0, \label{Eq_Npol} \\
&N_{g_zg_z'}=q^2\epsilon_{g_z-g_z'}-\sum_{g_z''}(k_z+g_z)\eta_{g_z-g_z''}(k_z+g_z'')\epsilon_{g_z''-g_z'}, 
\end{align}
and the electric displacement is obtained by Ampere's law $\nabla\times {\bm H}=-i\omega {\bm D}$ with ${\bm H}={\bm B}/\mu_0$.   
In the same way, by diagonalizing the second term in Eq. (\ref{Eq_Npol}), we obtain 
\begin{align}
&\psi_{lg_z}^N(\rho)=\sum_\alpha c_{l\alpha}^N J_l(\sigma_\alpha^N\rho) \psi_{\alpha g_z}^N,\label{Eq_Npolexp}\\ 
&\sum_{g_z'}N_{g_zg_z'}\psi_{\alpha g_z'}^N=(\sigma_\alpha^N)^2 \psi_{\alpha g_z}^N
\end{align}
Then, the radiation field inside the DBR pillar is expanded as a superposition of the $M$ and $N$ polarized eigenstates.

The boundary condition of the radiation field at $\rho=r_0$ is the continuity of the tangential components of ${\bm E}$ and ${\bm H}$ fields between the regions inside and outside the DBR pillar, and is given by  
\begin{widetext}
\begin{align}
&\sum_\alpha\left[ 
-\frac{\sigma_\alpha^M}{\epsilon_0} c_{l\alpha}^{M}J_l'(\sigma_\alpha^M r_0)
\tilde{\psi}_{\alpha g_z}^M 
-\frac{ilc^2}{\omega r_0} c_{l\alpha}^{N}J_l(\sigma_\alpha^N r_0)
\sum_{g_z'}\eta_{g_z-g_z'}(k_z+g_z')\psi_{\alpha g_z'}^N \right]\nonumber \\
&\hskip50pt =-\psi_{lg_z}^{M0}J_l'(\lambda_{g_z}r_0) -\psi_{lg_z}^{M{\rm ind}}H_l'(\lambda_{g_z}r_0) -\frac{l(k_z+g_z)}{q\lambda_{g_z}r_0}\left(\psi_{lg_z}^{N0}J_l(\lambda_{g_z}r_0) +\psi_{lg_z}^{N{\rm ind}}H_l(\lambda_{g_z}r_0) \right),\\
&i\frac{c^2}{\omega} \sum_\alpha c_{l\alpha}^{N} (\sigma_\alpha^N)^2 J_l(\sigma_\alpha^N r_0)\sum_{g_z'}\eta_{g_z-g_z'}\psi_{\alpha g_z'}^N =\frac{\lambda_{g_z}}{q}( \psi_{lg_z}^{N0}J_l(\lambda_{g_z}r_0) +\psi_{lg_z}^{N{\rm ind}}H_l(\lambda_{g_z}r_0) ),\\
&\sum_\alpha\left[ 
-\frac{\sigma_\alpha^N}{\mu_0} c_{l\alpha}^{N}J_l'(\sigma_\alpha^N r_0)
\psi_{\alpha g_z}^N 
+i\frac{lc^2(k_z+g_z)}{\omega r_0} c_{l\alpha}^{M}J_l(\sigma_\alpha^M r_0)
\tilde{\psi}_{\alpha g_z}^M \right]\nonumber \\
&\hskip50pt =\frac{1}{ic}\left[
-\psi_{lg_z}^{N0}J_l'(\lambda_{g_z}r_0) -\psi_{lg_z}^{N{\rm ind}}H_l'(\lambda_{g_z}r_0) -\frac{l(k_z+g_z)}{q\lambda_{g_z}r_0}\left(\psi_{lg_z}^{M0}J_l(\lambda_{g_z}r_0) +\psi_{lg_z}^{M{\rm ind}}H_l(\lambda_{g_z}r_0) \right)\right],\\
&\frac{c^2}{i\omega} \sum_\alpha c_{l\alpha}^{M} (\sigma_\alpha^M)^2 J_l(\sigma_\alpha^M r_0)\tilde{\psi}_{\alpha g_z}^M =\frac{\lambda_{g_z}}{i\omega}( \psi_{lg_z}^{M0}J_l(\lambda_{g_z}r_0) +\psi_{lg_z}^{M{\rm ind}}H_l(\lambda_{g_z}r_0) ). 
\end{align}
\end{widetext}
By eliminating $c_{l\alpha}^\beta$ numerically, we obtain the $t$-matrix defined in Eq. (\ref{Eq_tmat}).

\subsection{Layer-by-layer basis}

In the layer-by-layer basis, we solve the Maxwell equation analytically in each layer of the DBR, and consider the field continuity at the layer boundaries. The boundary condition in the radial direction of the DBR pillar is considered later.  The consistency with the Bloch theorem results in a secular equation for the possible eigenmodes  in the DBR.

In this basis, the so-called $P$ and $S$ polarizations are decoupled. 
In the $P$ polarization, the electric field in two adjacent layers (see Fig. \ref{Fig_fibgra}) is given by 
\begin{align}
&{\bm E}({\bm x})= \left\{ \begin{array}{l}
a_P^+ {\bm P}_a^+ {\rm e}^{{\rm i}{\bm K}_a^+\cdot({\bm x}-z_a\hat{z})} +a_P^- {\bm P}_a^- {\rm e}^{{\rm i}{\bm K}_a^-\cdot({\bm x}-z_a\hat{z})}   \\
\hskip100pt {\rm for}\quad |z-z_a|<\frac{d_a}{2} \\
b_P^+ {\bm P}_b^+ {\rm e}^{{\rm i}{\bm K}_b^+\cdot({\bm x}-z_b\hat{z})} +b_P^- {\bm P}_b^- {\rm e}^{{\rm i}{\bm K}_b^-\cdot({\bm x}-z_b\hat{z})} \\
 \hskip100pt {\rm for}\quad |z-z_b|<\frac{d_b}{2}
\end{array}\right. \\
& {\bm P}_\xi^\pm = \pm \frac{\gamma_\xi}{q_\xi}\hat{\bm k}_\| -\frac{|{\bm k}_\||}{q_\xi}\hat{z}, \quad {\bm K}_\xi^\pm = {\bm k}_\| \pm \gamma_\xi\hat{z},\\
&\gamma_\xi=\sqrt{q_\xi^2-{\bm k}_\|^2},\quad q_\xi=q\sqrt{\epsilon_\xi} \quad (\xi=a,b),\\
&{\bm k}_\|=(k_x,k_y),\quad \hat{\bm k}_\|=\frac{1}{|{\bm k}_\||}(k_x,k_y). 
\end{align}
Similarly, in the $S$ polarization, the electric field becomes 
\begin{align}
&{\bm E}({\bm x})= \left\{\begin{array}{l}
{\bm S} ( a_S^+ {\rm e}^{{\rm i}{\bm K}_a^+\cdot({\bm x}-z_a\hat{z})}                  + a_S^- {\rm e}^{{\rm i}{\bm K}_a^-\cdot({\bm x}-z_a\hat{z})} )\\
\hskip100pt {\rm for}\quad |z-z_a|<\frac{d_a}{2} \\
{\bm S} ( b_S^+ {\rm e}^{{\rm i}{\bm K}_b^+\cdot({\bm x}-z_b\hat{z})}                  + b_S^- {\rm e}^{{\rm i}{\bm K}_b^-\cdot({\bm x}-z_b\hat{z})} )\\
\hskip100pt {\rm for}\quad |z-z_b|<\frac{d_b}{2} \\
\end{array}\right.\\
&{\bm S}=\hat{\bm k}_\perp=\frac{1}{|{\bm k}_\||}(-k_y,k_x).    
\end{align}

 By imposing the continuity in the tangential component of ${\bm E}$ and ${\bm H}$ together with the Bloch theorem, 
the secular equation of the eigenmodes in the DBR becomes 
\begin{align}
&{\rm e}^{{\rm i}k_zd}=f_\sigma (k_\|) \pm \sqrt{(f_\sigma (k_\|))^2 -1} \quad (\sigma=P,S),\label{Eq_secular_DBR}\\ 
&f_\sigma (k_\|)=\frac{1}{2}\left(
{\rm e}^{{\rm i}\gamma_b d}\left(t_\sigma-\frac{r_\sigma^2}{t_\sigma}\right) 
+ {\rm e}^{-{\rm i}\gamma_b d} \frac{1}{t_\sigma}
\right),\\
&t_P=\frac{
{\rm e}^{{\rm i}(\gamma_a-\gamma_b)d_a}\frac{4\gamma_a\gamma_b}{\epsilon_a\epsilon_b} }
{ \left( \frac{\gamma_a}{\epsilon_a}+  \frac{\gamma_b}{\epsilon_b} \right)^2 -{\rm e}^{2{\rm i}\gamma_ad_a} \left( \frac{\gamma_a}{\epsilon_a}-  \frac{\gamma_b}{\epsilon_b} \right)^2   },\\
&r_P=\frac{  
\left(\left(\frac{\gamma_a}{\epsilon_a}\right)^2 -\left( \frac{\gamma_b}{\epsilon_b} \right)^2 \right) {\rm e}^{-{\rm i}\gamma_b d_a} ( {\rm e}^{2{\rm i}\gamma_ad_a} -1  )  }
{ \left( \frac{\gamma_a}{\epsilon_a}+  \frac{\gamma_b}{\epsilon_b} \right)^2 -{\rm e}^{2{\rm i}\gamma_ad_a} \left( \frac{\gamma_a}{\epsilon_a}-  \frac{\gamma_b}{\epsilon_b} \right)^2   },\\
&t_S=\frac{
{\rm e}^{{\rm i}(\gamma_a-\gamma_b)d_a}4\gamma_a\gamma_b }
{ \left( \gamma_a+ \gamma_b\right)^2 -{\rm e}^{2{\rm i}\gamma_ad_a} \left( \gamma_a - \gamma_b \right)^2   },\\
&r_S=\frac{  
\left(\gamma_a^2 -\gamma_b^2 \right) {\rm e}^{-{\rm i}\gamma_b d_a} ( {\rm e}^{2{\rm i}\gamma_ad_a} -1  )  }
{ \left( \gamma_a + \gamma_b \right)^2 -{\rm e}^{2{\rm i}\gamma_ad_a} \left( \gamma_a-  \gamma_b \right)^2   }. 
\end{align}

We have infinite number of solutions $k_\|^{\sigma\alpha}$ ($\alpha=1,2,\dots$) for Eq. (\ref{Eq_secular_DBR}), in which most solutions are evanescent, namely, $k_\|^{\sigma\alpha}$ is not real.  
The field coefficients ($a_\sigma^\pm$,$b_\sigma^\pm$) of mode $\alpha$ satisfy 
\begin{align}
&a_{\sigma\alpha}^+= \frac{1}{1-L_{\sigma\alpha}^{+-}R_{\sigma\alpha}^{-+}}(L_{\sigma\alpha}^{++}b_{\sigma\alpha}^+ + L_{\sigma\alpha}^{+-}R_{\sigma\alpha}^{--}
{\rm e}^{{\rm i}k_zd}b_{\sigma\alpha}^-),\\
&a_{\sigma\alpha}^-= \frac{1}{1-R_{\sigma\alpha}^{-+}L_{\sigma\alpha}^{+-}}(R_{\sigma\alpha}^{--}{\rm e}^{{\rm i}k_zd} b_{\sigma\alpha}^- + R_{\sigma\alpha}^{-+}L_{\sigma\alpha}^{++}b_{\sigma\alpha}^+),\\
&b_{\sigma\alpha}^+=-\frac{T_{\sigma\alpha}^{+-}}{ |T_{\sigma\alpha}^{++}|^2+
|T_{\sigma\alpha}^{+-}|^2 }{\rm e}^{-{\rm i}\gamma_b^{\sigma\alpha}d/2},\\
&b_{\sigma\alpha}^-=\frac{T_{\sigma\alpha}^{++}}{ |T_{\sigma\alpha}^{++}|^2+
|T_{\sigma\alpha}^{+-}|^2 } {\rm e}^{{\rm i}\gamma_b^{\sigma\alpha}d/2},\\
&T_{\sigma\alpha}^{++}={\rm e}^{{\rm i}\gamma_b^{\sigma\alpha}d}\left(t_{\sigma\alpha}-\frac{r_{\sigma\alpha}^2}{t_{\sigma\alpha}}\right)-{\rm e}^{{\rm i}k_zd},\\
&T_{\sigma\alpha}^{+-}={\rm e}^{{\rm i}\gamma_b^{\sigma\alpha}d}\frac{r_{\sigma\alpha}}{t_{\sigma\alpha}},\\
&L_{P\alpha}^{++} = \frac{2\frac{\gamma_b^{p\alpha}}{\sqrt{\epsilon_a\epsilon_b}}}{ \frac{\gamma_a^{p\alpha}}{\epsilon_a} +  \frac{\gamma_b^{p\alpha}}{\epsilon_b}}
{\rm e}^{{\rm i}(\gamma_a^{p\alpha}d_a + \gamma_b^{p\alpha}d_b)/2 },\\
&L_{P\alpha}^{+-} = \frac{\frac{\gamma_a^{p\alpha}}{\epsilon_a} -  \frac{\gamma_b^{p\alpha}}{\epsilon_b} }
{ \frac{\gamma_a^{p\alpha}}{\epsilon_a} +  \frac{\gamma_b^{p\alpha}}{\epsilon_b}}
{\rm e}^{{\rm i}\gamma_a^{p\alpha}d_a },\\
&L_{S\alpha}^{++} = \frac{2\gamma_b^{s\alpha}}{ \gamma_a^{s\alpha} + \gamma_b^{s\alpha}}
{\rm e}^{{\rm i}(\gamma_a^{s\alpha}d_a + \gamma_b^{s\alpha}d_b)/2 },\\
&L_{S\alpha}^{+-} = \frac{\gamma_a^{s\alpha} - \gamma_b^{s\alpha} }
{ \gamma_a^{s\alpha} +  \gamma_b^{s\alpha}}
{\rm e}^{{\rm i}\gamma_a^{s\alpha}d_a },\\
&R_{\sigma\alpha}^{--}=L_{\sigma\alpha}^{++},\quad 
R_{\sigma\alpha}^{-+}=L_{\sigma\alpha}^{+-},\\
&\gamma_\xi^{\sigma\alpha}=\sqrt{q_\xi^2-(k_\|^{\sigma\alpha})^2},\\
&t_{\sigma\alpha}=t_\sigma(k_\|^{\sigma\alpha}), \quad r_{\sigma\alpha}=r_\sigma(k_\|^{\sigma\alpha}). 
\end{align}

Let us consider an orthogonality of the eigenmodes inside the DBR. 
In terms of the Fourier modal basis, the $P$ polarization eigenstates satisfy 
\begin{align}
&\sum_{g_z'}\left(q^2\delta_{g_zg_z'}-(k_z+g_z)\eta_{g_z-g_z'}(k_z+g_z')\right)({\bm H}_{g_z'}^{(\alpha)})_\| \nonumber \\
&\hskip50pt =(k_\|^{P\alpha})^2  \sum_{g_z'}\eta_{g_z-g_z'}({\bm H}_{g_z'}^{(\alpha)})_\|.
\end{align}
Here, $k_\|^{P\alpha}$ corresponds to $\sigma_\alpha^N$ of the Fourier modal basis. 
If the DBR is lossless, we have a hermitian matrix of $[\eta]_{g_zg_z'}\equiv\eta_{g_z-g_z'}$. In this case, two eigenstates having different eigenvalues $(k_\|^{P\alpha})^2$ are orthogonal to each other as
\begin{align}
\sum_{g_zg_z'} \eta_{g_z-g_z'} ({\bm H}_{g_z}^{(\alpha)})_\|^* \cdot ({\bm H}_{g_z'}^{(\alpha')})_\| \propto \delta_{\alpha\alpha'}. 
\end{align} 
Similarly, the $S$-polarization eigenstates satisfy 
\begin{align}
\sum_{g_z'}M_{g_zg_z'}({\bm E}_{g_z'}^{(\alpha)})_\|=(k_\|^{S\alpha})^2 ({\bm E}_{g_z}^{(\alpha)})_\|, 
\end{align}
where $k_\|^{S\alpha}$ corresponds to $\sigma_\alpha^M$.  
Provided that the system is lossless, the orthogonality holds as 
\begin{align}
\sum_{g_z} ({\bm E}_{g_z}^{(\alpha)})_\|^* \cdot ({\bm E}_{g_z}^{(\alpha')})_\| \propto \delta_{\alpha\alpha'}. 
\end{align} 
This orthogonality is nontrivial in the layer-by-layer basis, but is translated as  
\begin{align}
&\frac{1}{d}\int_{\rm UC}{\rm d}z \psi_{P\alpha}^*(z) \psi_{P\alpha'}(z)\propto \delta_{\alpha\alpha'},\\
&\frac{1}{d}\int_{\rm UC}{\rm d}z \psi_{S\alpha}^*(z) \psi_{S\alpha'}(z)\propto \delta_{\alpha\alpha'},\\
&\psi_{\sigma\alpha}(z)=\left\{
\begin{array}{l}
a_{\sigma\alpha}^+ {\rm e}^{{\rm i}\gamma_a^{\sigma\alpha} (z-z_a)} + a_{\sigma\alpha}^- {\rm e}^{-{\rm i}\gamma_a^{\sigma\alpha} (z-z_a)} \\
\hskip100pt {\rm for} \quad |z-z_a|<\frac{d_a}{2} \\
b_{\sigma\alpha}^+ {\rm e}^{{\rm i}\gamma_b^{\sigma\alpha} (z-z_b)} + b_{\sigma\alpha}^- {\rm e}^{-{\rm i}\gamma_b^{\sigma\alpha} (z-z_b)} \\
\hskip100pt {\rm for} \quad |z-z_b|<\frac{d_b}{2} 
\end{array} \right., 
\end{align}
where a unit cell (UC) is taken to be $z_b-d_b/2\le z < z_a+d_a/2$.

Using the Bessel-function expansion of the plane wave, 
\begin{align}
{\rm e}^{{\rm i}{\bm k}_\|\cdot{\bm x}_\|}=\sum_l {\rm i}^l J_l(k_\|\rho){\rm e}^{{\rm i}l(\phi-\phi_{{\bm k}_\|})}, 
\end{align}
the $P$- and $S$-polarization eigenstates are written as 
\begin{align}
&{\bm E}^{P\alpha}({\bm x})= \sum_l {\rm i}^l J_l(k_\|^{P\alpha}\rho){\rm e}^{{\rm i}l(\phi-\phi_{{\bm k}_\|})} \nonumber \\
&\hskip20pt \times \left\{ \begin{array}{l}
a_{P\alpha}^+{\bm P}_a^{\alpha +}{\rm e}^{{\rm i}\gamma_a^{P\alpha} (z-z_a)} + a_{P\alpha}^-{\bm P}_a^{\alpha -}{\rm e}^{-{\rm i}\gamma_a^{P\alpha} (z-z_a)} \\
 \hskip80pt {\rm for} \quad |z-z_a|<\frac{d_a}{2}\\
b_{P\alpha}^+{\bm P}_b^{\alpha +}{\rm e}^{{\rm i}\gamma_b^{P\alpha} (z-z_b)} + b_{P\alpha}^-{\bm P}_b^{\alpha -}{\rm e}^{-{\rm i}\gamma_b^{P\alpha} (z-z_b)} \\
 \hskip80pt {\rm for} \quad |z-z_b|<\frac{d_b}{2} 
\end{array}\right.,\\
&{\bm E}^{S\alpha}({\bm x})= \sum_l {\rm i}^l J_l(k_\|^{S\alpha}\rho){\rm e}^{{\rm i}l(\phi-\phi_{{\bm k}_\|})} \nonumber \\
&\hskip20pt \times \left\{ \begin{array}{l}
a_{S\alpha}^+{\bm S}{\rm e}^{{\rm i}\gamma_a^{S\alpha} (z-z_a)} + a_{S\alpha}^-{\bm S}{\rm e}^{-{\rm i}\gamma_a^{S\alpha} (z-z_a)} \\
 \hskip80pt {\rm for} \quad |z-z_a|<\frac{d_a}{2}\\
b_{S\alpha}^+{\bm S}{\rm e}^{{\rm i}\gamma_b^{S\alpha} (z-z_b)} + b_{S\alpha}^-{\bm S}{\rm e}^{-{\rm i}\gamma_b^{S\alpha} (z-z_b)} \\
 \hskip80pt {\rm for} \quad |z-z_b|<\frac{d_b}{2} 
\end{array}\right..
\end{align}
Since the eigenvalues of the DBR are free from the orientation of ${\bm k}_\|$, we introduce angular-momentum eigenstates by 
\begin{align}
{\bm E}_l^{\sigma\alpha}({\bm x})=\int \frac{{\rm d}\phi_{{\bm k}_\|}}{2\pi} {\rm e}^{{\rm i}l\phi_{{\bm k}_\|}} {\bm E}^{\sigma\alpha}({\bm x}). 
\end{align}
Then, we expand the radiation field inside the DBR pillar by this basis:
\begin{align}
{\bm E}({\bm x})=\sum_{l\alpha}(C_{l}^{P\alpha}{\bm E}_l^{P\alpha}({\bm x})+ 
   C_{l}^{S\alpha}{\bm E}_l^{S\alpha}({\bm x}) ). \label{Eq_eigmode_exp} 
\end{align}

The boundary condition of the radiation field at the pillar surface ($\rho=r_0$) reads 
\begin{widetext}
\begin{align}
&\sum_{\alpha}\left[C_l^{P\alpha}{\rm i}^l \frac{l}{k_\|^{P\alpha}r_0}J_l(k_\|^{P\alpha}r_0)\psi_{P\alpha}^{E_\phi}(z) + C_{l}^{S\alpha} i^{l-1}J_l'(k_\|^{S\alpha}r_0)\psi_{S\alpha}(z) \right] \nonumber \\
& \hskip5pt = \sum_{g_z} \left[ -\psi_{lg_z}^{M0}J_l'(\lambda_{g_z}r_0) - \psi_{lg_z}^{M{\rm ind}}H_l'(\lambda_{g_z}r_0) - \frac{l(k_z+g_z)}{\lambda_{g_z}r_0q}(\psi_{lg_z}^{N0}J_l(\lambda_{g_z}r_0) + \psi_{lg_z}^{N{\rm ind}}H_l(\lambda_{g_z}r_0)) \right] {\rm e}^{{\rm i}(k_z+g_z)z},\label{Eq_Ep_layer}\\
&-\sum_\alpha C_l^{P\alpha}{\rm i}^l J_l(k_\|^{P\alpha}r_0)\psi_{P\alpha}^{E_z}(z)
=\sum_{g_z} \frac{\lambda_{g_z}}{q}( \psi_{lg_z}^{N0}J_l(\lambda_{g_z}r_0) +\psi_{lg_z}^{N{\rm ind}}H_l(\lambda_{g_z}r_0) ){\rm e}^{{\rm i}(k_z+g_z)z},\label{Eq_Ez_layer}\\
&\sum_{\alpha}\left[-C_l^{S\alpha}{\rm i}^l \frac{l}{k_\|^{S\alpha}r_0}J_l(k_\|^{S\alpha}r_0)\psi_{S\alpha}^{H_\phi}(z) + C_{l}^{P\alpha} i^{l-1}J_l'(k_\|^{P\alpha}r_0)\sqrt{\epsilon(z)}\psi_{P\alpha}(z) \right] \nonumber \\
& \hskip5pt = -i\sum_{g_z} \left[ -\psi_{lg_z}^{N0}J_l'(\lambda_{g_z}r_0) - \psi_{lg_z}^{N{\rm ind}}H_l'(\lambda_{g_z}r_0) - \frac{l(k_z+g_z)}{\lambda_{g_z}r_0q}(\psi_{lg_z}^{M0}J_l(\lambda_{g_z}r_0) + \psi_{lg_z}^{M{\rm ind}}H_l(\lambda_{g_z}r_0)) \right] {\rm e}^{{\rm i}(k_z+g_z)z},\label{Eq_Hp_layer}\\
&\sum_\alpha C_l^{S\alpha}{\rm i}^l J_l(k_\|^{S\alpha}r_0)k_\|^{S\alpha}\psi_{S\alpha}(z)
=-i\sum_{g_z} \lambda_{g_z} \left(\psi_{lg_z}^{M0}J_l(\lambda_{g_z}r_0) +\psi_{lg_z}^{M{\rm ind}}H_l(\lambda_{g_z}r_0) \right){\rm e}^{{\rm i}(k_z+g_z)z},\label{Eq_Hz_layer}
\end{align} 
\end{widetext} 
where 
\clearpage
\begin{align}
&\psi_{P\alpha}^{E_\phi}(z)=\left\{\begin{array}{l}
\frac{\gamma_a^{P\alpha}}{q_a}(a_{P\alpha}^+ {\rm e}^{{\rm i}\gamma_a^{P\alpha}(z-z_a)} - a_{P\alpha}^- {\rm e}^{-{\rm i}\gamma_a^{P\alpha}(z-z_a)})    \\
\hskip80pt {\rm for} \quad |z-z_a|<\frac{d_a}{2} \\
\frac{\gamma_b^{P\alpha}}{q_b}(b_{P\alpha}^+ {\rm e}^{{\rm i}\gamma_b^{P\alpha}(z-z_b)} - b_{P\alpha}^- {\rm e}^{-{\rm i}\gamma_b^{P\alpha}(z-z_b)}) \\
\hskip80pt {\rm for} \quad |z-z_b|<\frac{d_b}{2} 
 \end{array}\right.,\\
&\psi_{P\alpha}^{E_z}(z)=\left\{\begin{array}{l}
\frac{k_\|^{P\alpha}}{q_a}(a_{P\alpha}^+ {\rm e}^{{\rm i}\gamma_a^{P\alpha}(z-z_a)} + a_{P\alpha}^- {\rm e}^{-{\rm i}\gamma_a^{P\alpha}(z-z_a)})    \\
\hskip80pt {\rm for} \quad |z-z_a|<\frac{d_a}{2} \\
\frac{k_\|^{P\alpha}}{q_b}(b_{P\alpha}^+ {\rm e}^{{\rm i}\gamma_b^{P\alpha}(z-z_b)} + b_{P\alpha}^- {\rm e}^{-{\rm i}\gamma_b^{P\alpha}(z-z_b)}) \\
\hskip80pt {\rm for} \quad |z-z_b|<\frac{d_b}{2} 
 \end{array}\right.,\\ 
&\psi_{S\alpha}^{H_\phi}(z)=\left\{\begin{array}{l}
\frac{\gamma_a^{S\alpha}}{q_a}(a_{S\alpha}^+ {\rm e}^{{\rm i}\gamma_a^{S\alpha}(z-z_a)} - a_{S\alpha}^- {\rm e}^{-{\rm i}\gamma_a^{S\alpha}(z-z_a)})    \\
\hskip80pt {\rm for} \quad |z-z_a|<\frac{d_a}{2} \\
\frac{\gamma_b^{S\alpha}}{q_b}(b_{S\alpha}^+ {\rm e}^{{\rm i}\gamma_b^{S\alpha}(z-z_b)} - b_{S\alpha}^- {\rm e}^{-{\rm i}\gamma_b^{S\alpha}(z-z_b)}) \\
\hskip80pt {\rm for} \quad |z-z_b|<\frac{d_b}{2} 
 \end{array}\right..
\end{align}
Here, we expand the right hand side of Eqs. (\ref{Eq_Ep_layer}) and (\ref{Eq_Hp_layer}) in terms of the $S$- and $P$-polarized eigenmodes  $\psi_{\sigma\alpha}(z)$, respectively,  and compare the eigenmode-expansion coefficients between the left hand side and right hand side of these equations. Therefore, we have $2N_{\rm e}$ equations, where $N_{\rm e}$ stands for the number of the eigenmodes for each polarization, taken into account in the numerical calculation. On the other hand, we Fourier-expand the left hand sides of Eqs. (\ref{Eq_Ez_layer}) and (\ref{Eq_Hz_layer}), and compare the  Fourier-expansion coefficients. We thus obtain $2N_{\rm r}$ equations, where $N_{\rm r}$ is the number of the reciprocal lattices $g_z$ taken into account in the numerical calculation.  
In this way we have $2(N_{\rm e}+N_{\rm r})$ linear equations for $2(N_{\rm e}+N_{\rm r})$ unknown coefficients ($C_l^{\sigma\alpha}$ and $\psi_{lg_z}^{\beta{\rm ind}}$) per each $l$, provided the vector-cylindrical-wave expansion coefficients  $\psi_{lg_z}^{\beta 0}$ of the incident light.  
The $t$-matrix is thus obtained.

\subsection{Core-shell DBR pillars}
So far, we have focused on a simple DBR pillar composed of alternating finite-height pillars with the circular cross section.  
In a core-shell DBR pillar, we have the DBR of the inner core capped by homogeneous medium as the outer shell. However, it is rather easy in formulation to generalize the homogeneous outer shell to another DBR structure with the same period as in the core. In this case, for the core region  we employ the same expansion with the Bessel function as in Eqs. (\ref{Eq_Mpolexp}) and (\ref{Eq_Npolexp}) for the Fourier modal basis or in Eq. (\ref{Eq_eigmode_exp}) for the layer-by-layer basis. However, for the outer shell we expand the radiation field with both the Bessel function and Hankel function of the first kind. By imposing the boundary condition of the continuity of the radiation field at the interface between the core and shell, we obtain the $t$-matrix of the core-shell DBR pillar.  As far as coaxial (concentric) core-shell structures are considered, the resulting $t$-matrix is still diagonal with respect to angular momentum index $l$.

\subsection{1D photonic band structure}

Once we have the $t$-matrix numerically, we can evaluate various optical responses of the DBR pillar. The most important information here is the 1D photonic band structure, from which we can understand many features of the optical responses. The 1D photonic bands  consist of true-guided modes outside the light cone $\omega < c|k_z|$ and quasi-guided modes inside the light cone $\omega > c|k_z|$.  The latter modes merge with the radiation continuum and thus have finite life times. 

To evaluate the band structure of the true-guided modes, we just need to solve 
\begin{align}
{\rm det}[t_l]^{-1}=0.   
\end{align}
For the quasi-guided modes, we first evaluate the optical density of states (DOS) as  a function of $\omega$ and $k_z$, and then follow its peaks. 
The DOS of the modes with angular momentum $l$ is given by  
\begin{align}
&\Delta \rho_l(\omega,k_z)=\frac{1}{2\pi}\frac{\partial}{\partial \omega} {\rm arg}({\rm det}[U_l]),\\ 
&[U_{l}]_{g_z^og_z^o{}'}^{\beta\beta'}=\delta_{g_z^og_z^o{}'}\delta_{\beta\beta'}+2[t_l]_{g_z^og_z^o{}'}^{\beta\beta'},
\end{align}
where $g_z^o$ represents open diffraction channels. Matrix $U_l$ defined above is unitary in lossless cases.  
Strictly speaking, $\Delta \rho_l(\omega,k_z)$ is the increment of the optical DOS (relative to the DOS in vacuum) due to the presence of the DBR pillar, at fixed $\omega$ and $k_z$.

Around the resonance frequency $\omega_c$ of a quasi-guided mode, the relevant determinant behaves like \cite{landau2013quantum}
\begin{align}
{\rm det}[U_l]\simeq\frac{\omega-\omega_c-{\rm i}\Gamma}{\omega-\omega_c+{\rm i}\Gamma}, 
\end{align}
which results in the Lorentzian form of the DOS:
\begin{align}
&\Delta \rho_l(\omega,k_z)\simeq  \frac{\frac{\Gamma}{\pi}}{(\omega-\omega_c)^2+\Gamma^2}.
\end{align}  
The quality factor $Q$ of the mode is simply given by $Q=\omega_c/(2\Gamma)$.

\begin{figure*}
\begin{center}
\centerline{
\includegraphics[width=0.45\textwidth]{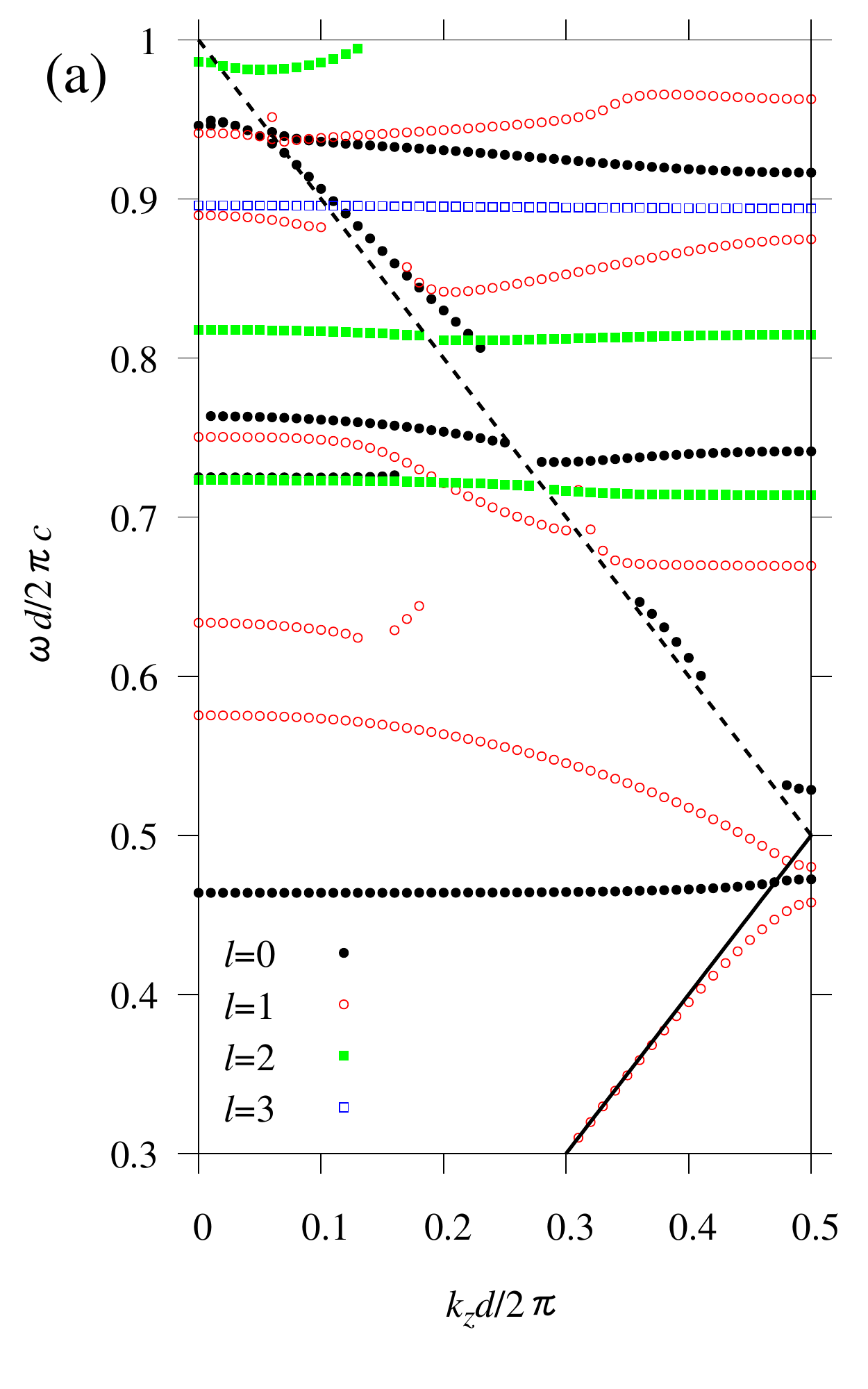}
\includegraphics[width=0.45\textwidth]{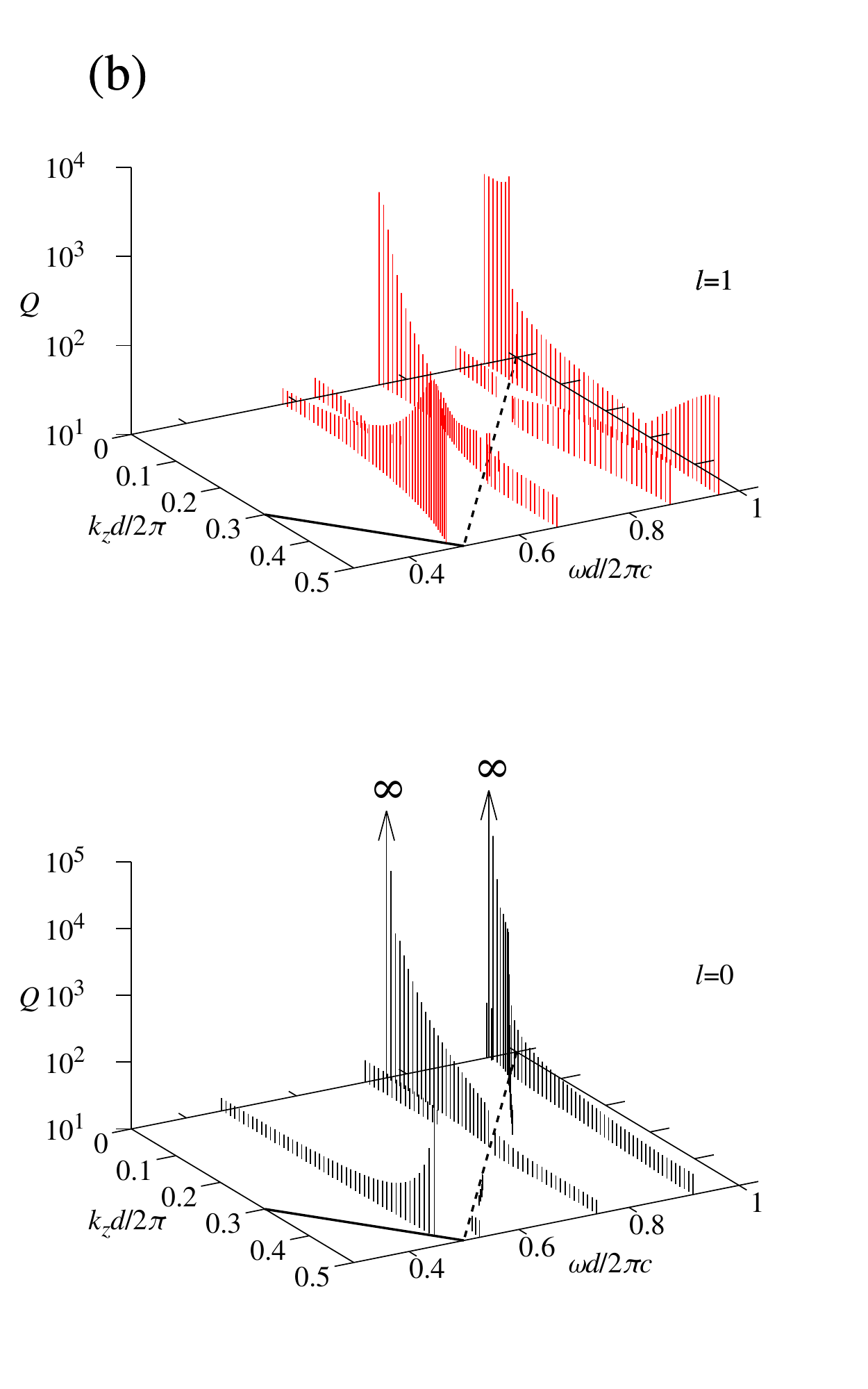}}
\end{center}
\caption{\label{Fig_band1d} (a) 1D photonic band structure of the DBR pillar. The photonic band modes are classified according to angular momentum $l$. (b) $Q$-values of the quasi-guided modes of $l=0$ and $l=1$. 
The DBR pillar has the following parameters (see Fig. \ref{Fig_fibgra}): 
$\epsilon_a=12$, $\epsilon_b=2$, $d_a=0.3d$, $d_b=0.7d$, $r_0=0.3d$.  Solid line is the light line $\omega=c|k_z|$. Dashed line represents the threshold of the Bragg diffraction.  
}
\end{figure*}

Figure \ref{Fig_band1d} shows the 1D photonic band structure and quality factors of a typical DBR pillar.
One can see clearly the photonic band structure formation both inside and outside the light cone. Outside the light cone $\omega <c|k_z|$, only a few modes of $l=0,\pm 1$ exist. The degenerate single-mode region of $l=\pm 1$ is $\omega d/2\pi c <0.45$. Inside the light cone, many bands of the quasi-guided modes are found.  Some of them have high $Q$ with well-defined dispersion curves, but some of them have low $Q$ having ill-defined dispersion curves terminated off the Brillouin-zone boundary.

It is remarkable The $Q$ value spectrum of $l=0$ exhibits the sharp peaks at $k_z=0$. In fact, the $Q$ value becomes infinity there for certain modes of $l=0$. This is the  so-called  bound states in the continuum \cite{hsu2016bound} which is caused by a symmetry mismatch. Such modes commonly have vanishing 
$\psi_{l=0;g_z=0}^{\beta{\rm ind}}$ components.

Let us consider why this happens. Possible continuum radiation modes that can couple to the eigenmodes of zero Bloch momentum $(k_z=0)$ are in-plane ($xy$ plane) propagating modes. 
The Bragg-diffraction channels of $g_z\ne 0$ are all evanescent provided $\omega <2\pi c/d$, so that we can neglect them. 
Suppose that the in-plane modes are propagating in $+y$ direction, for instance.  
They are given by a superposition of ${\bm E}=\hat{z}\exp(iqy)$ and $\hat{x}\exp(iqy)$. These basis have $(\sigma_x,\sigma_z)=(1,-1)$ and $(\sigma_x,\sigma_z)=(-1,1)$, respectively, where $\sigma_{x(z)}$ is the parity eigenvalue in the $x(z)$ direction. 
Besides, the eigenmodes of $l=0$ at $k_z=0$ in the DBR pillar are also classified by the parities. If an eigenmode has $(\sigma_x,\sigma_z)=(1,1)$ or (-1,-1), it cannot couple to the external propagating radiation because of the symmetry mismatch. Therefore, it has infinite $Q$.  
The parity constraint on $\psi_{l=0;g_z}^{\beta{\rm ind}}$ is as follows:
\begin{align}
-\psi_{l=0;g_z}^{M{\rm ind}}&=\sigma_x  \psi_{l=0;g_z}^{M{\rm ind}},\\
 \psi_{l=0;g_z}^{N{\rm ind}}&=\sigma_x  \psi_{l=0;g_z}^{N{\rm ind}},\\
 \psi_{l=0;-g_z}^{M{\rm ind}}&=\sigma_z  \psi_{l=0;g_z}^{M{\rm ind}},\\
-\psi_{l=0;-g_z}^{N{\rm ind}}&=\sigma_z  \psi_{l=0;g_z}^{N{\rm ind}}.
\end{align} 
Therefore, if an eigenmode has $(\sigma_x,\sigma_z)=(1,1)$ or (-1,-1), it has vanishing $\psi_{l=0;g_z=0}^{M{\rm ind}}$ and  $\psi_{l=0;g_z=0}^{N{\rm ind}}$. 
This is the case happened for the quasi-guided modes of $Q=\infty$ in Fig. \ref{Fig_band1d} (b).

We can also find peaks of the $Q$ value at genetic points in the Brillouin zone. This is a precursor of the bound states in the continuum not caused by a symmetry mismatch. Such phenomena are predicted for DBR pillars by Bulgakov and Sadreev \cite{PhysRevA.96.013841} and Gao et al \cite{gao2017bound}.

The presence of high $Q$ quasi-guided modes inside the light cone has a strong influence on light scattering by the DBR pillars. Let us consider the scattering cross section by the DBR pillar. The (elastic) scattering cross section $\sigma_{\rm cs}$ of the DBR pillar is given by  
\begin{align}
&\sigma_{\rm cs}=\sum_{g_z^o}\int_0^{2\pi} d\phi|{\bm f}_{g_z^o}(\phi)|^2,\\  
&{\bm f}_{g_z^o}(\phi)=\sqrt{\frac{2}{\pi\lambda_{g_z^o}}}
\sum_{l,\beta}(-{\rm i})^{l+1}{\rm e}^{{\rm i}l\phi}[{\bm P}_{g_z^o}^\beta]_{ll'}\psi_{l'g_z^o}^{\beta,{\rm ind}},
\end{align}
where ${\bm f}_{g_z^o}(\phi)$ describes the far-field pattern of the induced radiation field: 
\begin{align}
{\bm E}^{\rm ind}({\bm x})\simeq \sum_{g_z^o}\frac{1}{\sqrt{\rho}} {\rm e}^{{\rm i}\lambda_{g_z^o}\rho+{\rm i}(k_z+g_z^o)z+{\rm i}\frac{\pi}{4}}{\bm f}_{g_z^o}(\phi).
\end{align} 
In a uniform pillar of infinite height, the scattering cross section approaches to the twice of the geometric cross section (diameter of the circular pillar) in the high-frequency limit.

\begin{figure*}
\begin{center}
\includegraphics[width=0.45\textwidth]{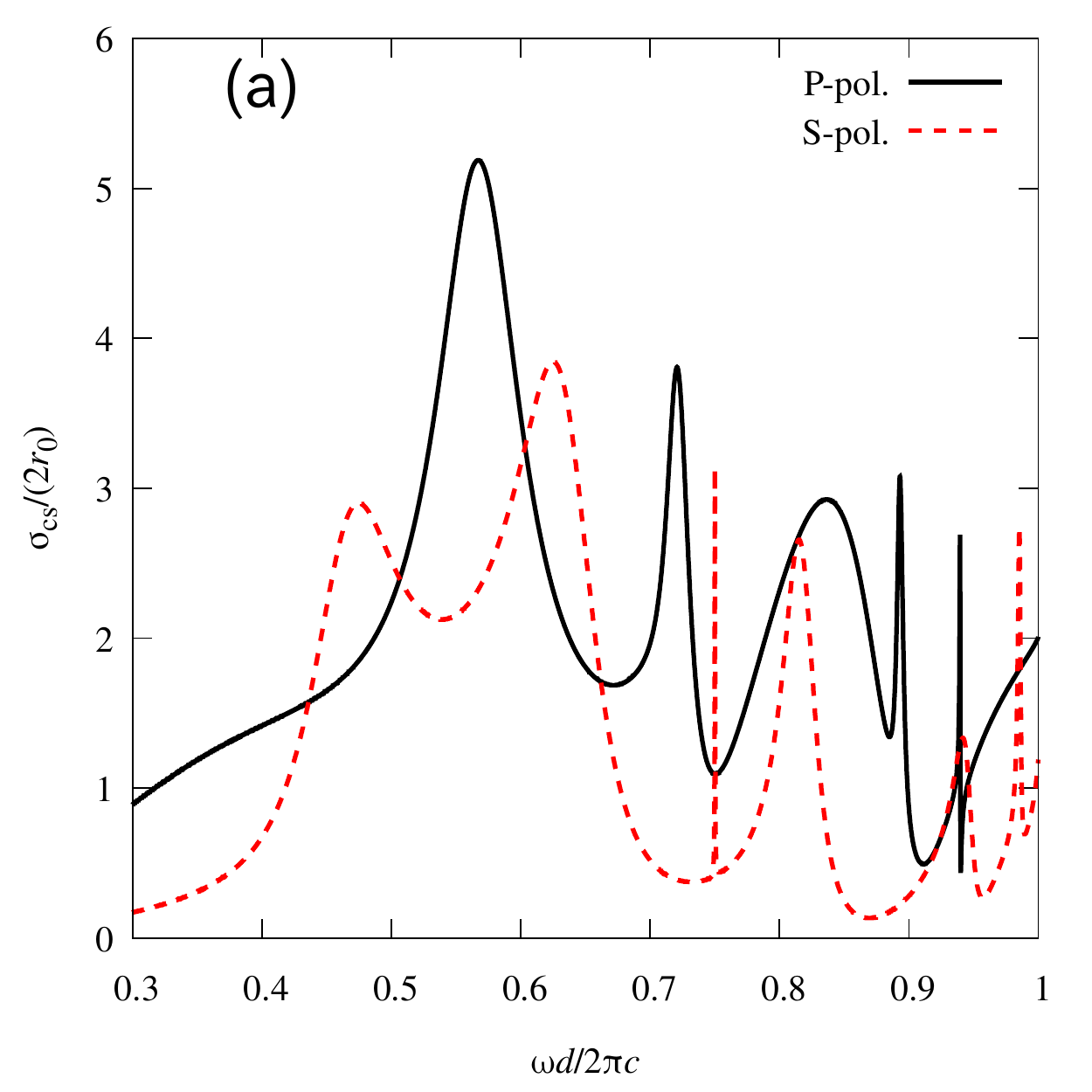}
\includegraphics[width=0.45\textwidth]{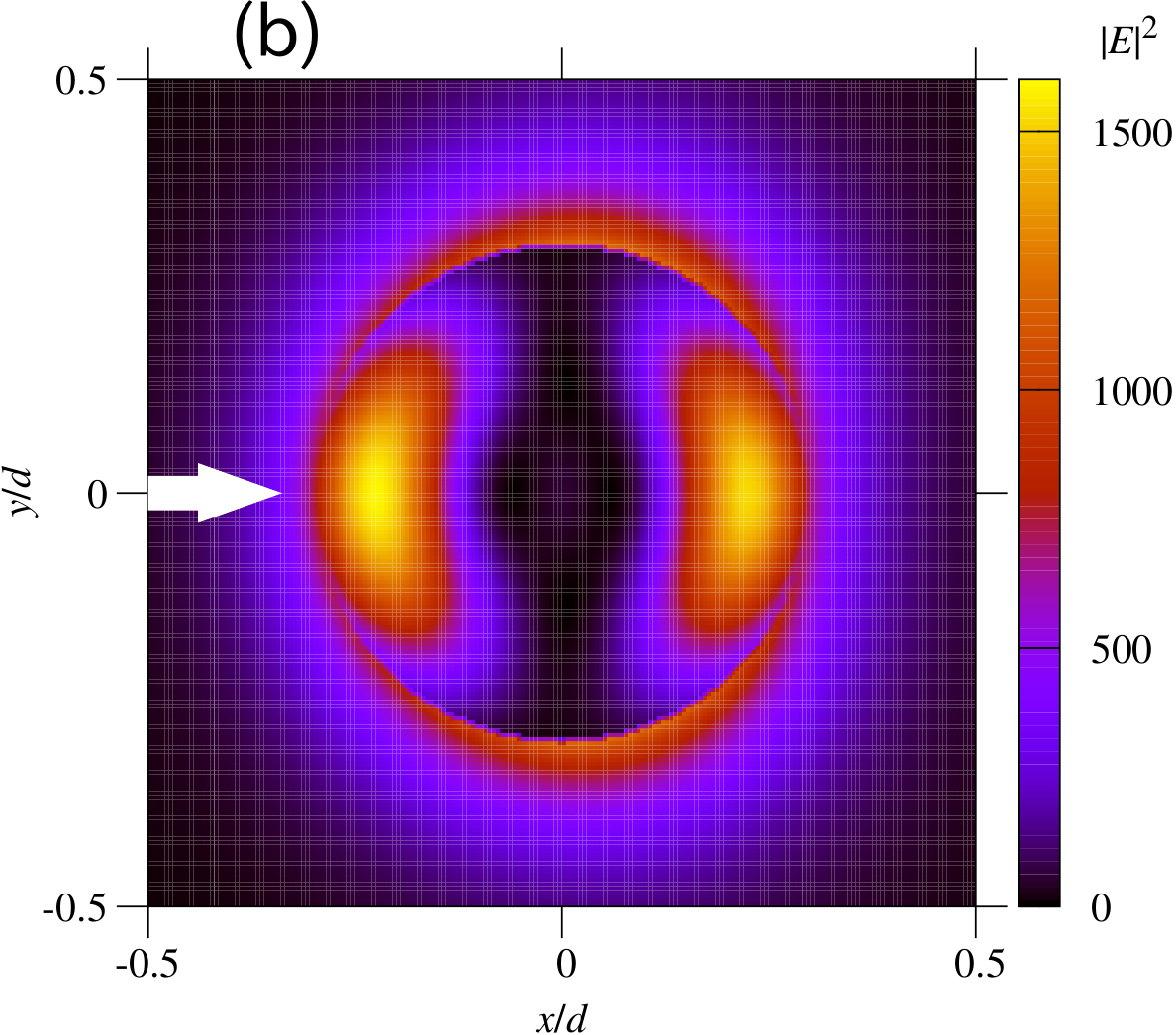}
\end{center}
\caption{\label{Fig_scatcs} (a) The scattering cross section of the DBR pillar 
for normally incident plane wave. Both P and S polarizations are considered. The  parameters are the same as in Fig. \ref{Fig_band1d}.   
(b) Electric-field intensity for the S-polarized incident light at $\omega d/2\pi c=0.75$. The incident light is coming from the left. The electric field intensity of the incident light is taken to be unity.  The field configuration on the plane bisecting the high-index pillar of $\epsilon_a=12$ is shown.    
}
\end{figure*}

Figure \ref{Fig_scatcs} (a) shows the scattering cross section as a function of frequency for the P and S-polarized incident light. 
The P-polarization stands for the electric-field polarization inside the incident plane (the plane formed by the incident wave vector and vector parallel to the pillar axis). The S-polarization is the polarization perpendicular to the incident plane.  The normal incidence ($k_z^0=0$) is assumed.  
The cross-section spectrum exhibits a sequence of peaks. 
As we can easily check, the peak positions corresponds to the photonic band modes at $k_z=0$ given in Fig. \ref{Fig_band1d} (a). If the mode concerned has high $Q$, the relevant peak becomes sharp. If the mode has infinite $Q$, however, the mode is invisible in the cross-section spectrum.  
The Electric-field configuration induced by the incident light of S-polarization at a sharp peak of $\omega d/2\pi c=0.75$  is shown in Fig. \ref{Fig_scatcs} (b). The Electric field is strongly enhanced inside the DBR pillar. Also, the dipole-like profile of $l=1$ is clearly visible.  The relevant mode at $\omega d/2\pi c=0.75$ in Fig. \ref{Fig_band1d} (a) certainly has $l=1$.

For oblique incidence, similar spectra with a sequence of peaks are obtained. 
In this case, the peak frequencies corresponds to the photonic band modes on the line of $k_z=(\omega/c)\sin\theta_{\rm inc}$ with incident angle $\theta_{\rm inc}$,  in Fig. \ref{Fig_band1d} (a). 
In this way, the 1d photonic band structure is very useful to understand  optical properties of the DBR pillar

\section{Array of DBR pillars}

It is straight-forward to apply the formalism to periodic and random arrangements of the DBR pillars. In such structures, further nontrivial phenomena of light transport and light-matter interaction can take place because of multiple scattering among the DBR pillars.  In a periodic case, the system can be regarded as a 2D or 3D PhC, so that it has a plenty of interesting phenomena in optics \cite{joannopoulos2011photonic}. 
In a random case, the Anderson localization of light becomes enriched.

To study such systems, the multiple-scattering formalism \cite{twersky1952multiple} is employed. 
Suppose that we have a parallel arrangement of DBR pillars with the same period $d$. The central coordinate and $t$-matrix of $\mu$-th DBR pillar ($\mu=1,2,...,N$) are denoted by $({\bm \rho}_\mu,z_\mu)$ and $[t_{\mu l}]_{g_zg_z'}^{\beta\beta'}$, respectively.  The incident radiation field to and induced radiation fields from the $\mu$-th DBR pillar are expressed as 
\begin{align}
&{\bm E}^{0}({\bm x})= \sum_{g_z,l,l',\beta}J_l(\lambda_{g_z}|{\bm \rho}-{\bm \rho}_\mu|){\rm e}^{{\rm i}l\phi({\bm \rho}-{\bm \rho}_\mu)}{\rm e}^{{\rm i}(k_z+g_z)(z-z_\mu)}\nonumber\\
&\hskip50pt \times [{\bm P}_{g_z}^{\beta}]_{ll'}\psi_{\mu l'g_z}^{\beta 0},\\
&{\bm E}_\mu^{\rm ind}({\bm x})= \sum_{g_z,l,l',\beta}H_l(\lambda_{g_z}|{\bm \rho}-{\bm \rho}_\mu|){\rm e}^{{\rm i}l\phi({\bm \rho}-{\bm \rho}_\mu)}{\rm e}^{{\rm i}(k_z+g_z)(z-z_\mu)}\nonumber\\
&\hskip50pt \times [{\bm P}_{g_z}^{\beta}]_{ll'}\psi_{\mu l'g_z}^{\beta {\rm ind}}.
\end{align}
The self-consistent equation for the induced radiation field via the multiple scattering  is given by
\begin{widetext}
\begin{align}
&\psi_{\mu lg_z}^{\beta {\rm ind}}=\sum_{g_z'\beta'}[t_{\mu l}]_{g_zg_z'}^{\beta\beta'}\left(
\psi_{\mu lg_z'}^{\beta' 0} + \sum_{l'}\sum_{\mu'(\ne\mu)}H_{l'-l}(\lambda_{g_z'}|{\bm \rho}_{\mu\mu'}|){\rm e}^{{\rm i}(l'-l)\phi({\bm \rho}_{\mu\mu'})}{\rm e}^{{\rm i}(k_z+g_z')z_{\mu\mu'}} \psi_{\mu' l'g_z'}^{\beta'{\rm ind}} \right),\\
&{\bm \rho}_{\mu\mu'}={\bm \rho}_{\mu}-{\bm \rho}_{\mu'},\quad 
z_{\mu\mu'}=z_{\mu}-z_{\mu'}. 
\end{align}
\end{widetext}
This equation becomes a foundation in the following part of the paper.

\subsection{1D periodic arrays}  

A 1D periodic array of DBR pillars forms a 2D PhC slab. In this system, the so-called layer-Korringa-Kohn-Rostoker (KKR) method \cite{Pendry-LEED-book,OHTAKA:N::73:p411-413:1979,leung1999computation} can be employed. The method allows us to study light transport in the 2D PhC slab composed of DBR pillars. Also, possible surface or planer-defect states in a 3D PhC composed of DBR pillars can be investigated with this formulation.  The key ingredient is the $S$-matrix of the 2D PhC slabs.

\begin{figure}
\begin{center}
\includegraphics[width=0.4\textwidth]{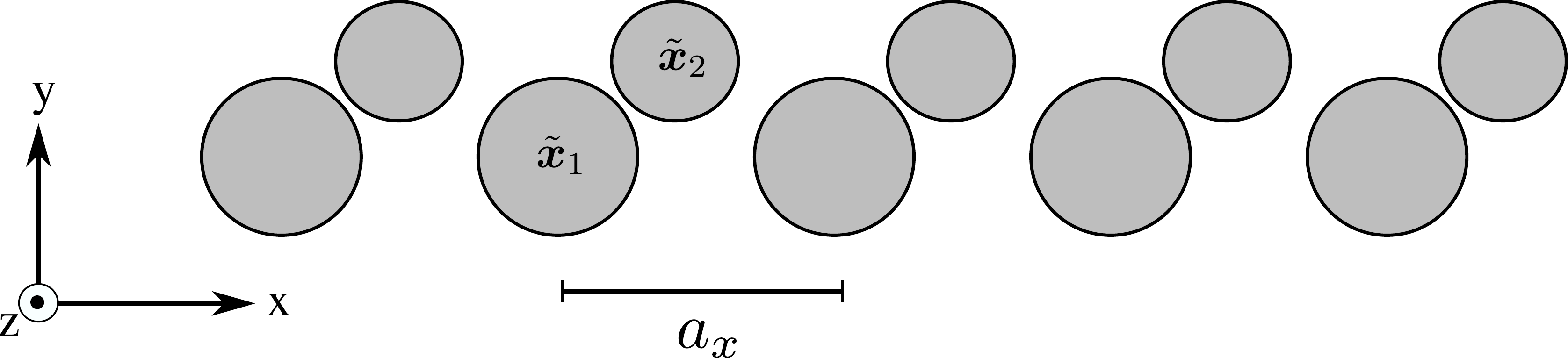}
\end{center}
\caption{\label{Fig_monolayer} Schematic illustration of the monolayer system under study (top view). It consists of a periodic arrangement of DBR pillars with lattice constant $a_x$. 
} 
\end{figure}

Let us consider the light scattering by a one-monolayer-thick PhC slab. 
A schematic illustration of the system under study is shown in Fig. \ref{Fig_monolayer}. 
The one-monolayer slab consists of a 1D periodic arrangement (in the $x$ direction) of the DBR pillars. The lattice constant is taken to be $a_x$. The central coordinate of the $\mu$-th DBR pillar is given by 
\begin{align}
{\bm x}_\mu=\tilde{\bm x}_s + n_xa_x \hat{x}\quad (n_x\in {\bm Z}),
\end{align}
where $s$ is the index that counts the number of the DBR pillars in a unit cell.

Suppose that an incident light is coming from both the upper and lower empty space in the  $y$ direction.  The incident electric field is generally written as 
\begin{align}
&{\bm E}^0({\bm x})=\sum_{\bm g}\left( 
{\rm e}^{{\rm i}{\bm K}_{\bm g}^+\cdot{\bm x}}{\bm p}_{\bm g}^+ + 
{\rm e}^{{\rm i}{\bm K}_{\bm g}^-\cdot{\bm x}}{\bm p}_{\bm g}^- \right), \\
&{\bm K}_{\bm g}^\pm =(k_x+g_x,\pm \gamma_{\bm g},k_z+g_z), \quad 
{\bm K}_{\bm g}^\pm \cdot {\bm p}_{\bm g}^\pm =0,\\
&\gamma_{\bm g}=\sqrt{q^2 - (k_x+g_x)^2-(k_z+g_z)^2}, \quad g_x=\frac{2\pi {\bm Z}}{a_x}.  
\end{align}
The induced electric field in the upper(+) and lower(-) empty space is written as
\begin{align}
&{\bm E}^{{\rm ind}\pm}({\bm x})=\sum_{\bm g}{\rm e}^{{\rm i}{\bm K}_{\bm g}^\pm \cdot{\bm x}} \sum_{{\bm g}'}\left( 
\tensor{S}_{{\bm g}{\bm g}'}^{\pm +} {\bm p}_{{\bm g}'}^+ + 
\tensor{S}_{{\bm g}{\bm g}'}^{\pm -} {\bm p}_{{\bm g}'}^- \right), 
\end{align} 
where $\tensor{S}_{{\bm g}{\bm g}'}^{\pm\pm}$ is the $S$-matrix.   
Using the multiple scattering formalism, the $S$-matrix is given by 
\begin{widetext}
\begin{align}
&\tensor{S}_{{\bm g}{\bm g}'}^{++}=\tensor{1}\delta_{{\bm g}{\bm g}'}
+\frac{2}{\gamma_{\bm g}a_x}\sum_{ss'll'\beta\beta'}{\rm e}^{-{\rm i}{\bm K}_{\bm g}^+\cdot\tilde{\bm x}_s}
(-{\rm i})^l {\rm e}^{{\rm i}l\phi({\bm K}_{\bm g}^+)}{\bm P}_{g_z}^{\beta}T_{ss';ll';g_zg_z'}^{\beta\beta'}\tilde{\bm P}_{g_z'}^{\beta'}{\rm i}^l {\rm e}^{-{\rm i}l\phi({\bm K}_{{\bm g}'}^+)}
{\rm e}^{{\rm i}{\bm K}_{{\bm g}'}^+\cdot\tilde{\bm x}_{s'}},\\
&\tensor{S}_{{\bm g}{\bm g}'}^{+-}=\hskip38pt 
\frac{2}{\gamma_{\bm g}a_x}\sum_{ss'll'\beta\beta'}{\rm e}^{-{\rm i}{\bm K}_{\bm g}^+\cdot\tilde{\bm x}_s}
(-{\rm i})^l {\rm e}^{{\rm i}l\phi({\bm K}_{\bm g}^+)}{\bm P}_{g_z}^{\beta}T_{ss';ll';g_zg_z'}^{\beta\beta'}\tilde{\bm P}_{g_z'}^{\beta'}{\rm i}^l {\rm e}^{-{\rm i}l\phi({\bm K}_{{\bm g}'}^-)}
{\rm e}^{{\rm i}{\bm K}_{{\bm g}'}^-\cdot\tilde{\bm x}_{s'}},\\
&\tensor{S}_{{\bm g}{\bm g}'}^{-+}=\hskip38pt
\frac{2}{\gamma_{\bm g}a_x}\sum_{ss'll'\beta\beta'}{\rm e}^{-{\rm i}{\bm K}_{\bm g}^-\cdot\tilde{\bm x}_s}
(-{\rm i})^l {\rm e}^{{\rm i}l\phi({\bm K}_{\bm g}^-)}{\bm P}_{g_z}^{\beta}T_{ss';ll';g_zg_z'}^{\beta\beta'}\tilde{\bm P}_{g_z'}^{\beta'}{\rm i}^l {\rm e}^{-{\rm i}l\phi({\bm K}_{{\bm g}'}^+)}
{\rm e}^{{\rm i}{\bm K}_{{\bm g}'}^+\cdot\tilde{\bm x}_{s'}},\\
&\tensor{S}_{{\bm g}{\bm g}'}^{--}= \tensor{1}\delta_{{\bm g}{\bm g}'}
+\frac{2}{\gamma_{\bm g}a_x}\sum_{ss'll'\beta\beta'}{\rm e}^{-{\rm i}{\bm K}_{\bm g}^-\cdot\tilde{\bm x}_s}
(-{\rm i})^l {\rm e}^{{\rm i}l\phi({\bm K}_{\bm g}^-)}{\bm P}_{g_z}^{\beta}T_{ss';ll';g_zg_z'}^{\beta\beta'}\tilde{\bm P}_{g_z'}^{\beta'}{\rm i}^l {\rm e}^{-{\rm i}l\phi({\bm K}_{{\bm g}'}^-)}
{\rm e}^{{\rm i}{\bm K}_{{\bm g}'}^-\cdot\tilde{\bm x}_{s'}},\\
&[T(k_x,k_z)]_{ss';ll';g_zg_z'}^{\beta\beta'}=\sum_{g_z''\beta''}[(1-t(k_z)G(k_x,k_z))^{-1}]_{ss';ll';g_zg_z''}^{\beta\beta''}[t_{s'l'}(k_z)]_{g_z''g_z'}^{\beta''\beta'},\\
&[1-t(k_z)G(k_x,k_z)]_{ss';ll';g_zg_z'}^{\beta\beta'}=\delta_{ss'}\delta_{ll'}\delta_{g_zg_z'}\delta_{\beta\beta'}- [t_{sl}(k_z)]_{g_zg_z'}^{\beta\beta'}[G(k_x,k_z)]_{ss';ll';g_z'}, \\
&[G(k_x,k_z)]_{ss';ll';g_z}=\sum_{n_x\in{\bm Z}}{}' {\rm e}^{{\rm i}k_xa_xn_x}H_{l'-l}(\lambda_{g_z}|\tilde{\bm \rho}_{ss'}-n_xa_x\hat{x}|){\rm e}^{{\rm i}(l'-l)\phi(\tilde{\bm \rho}_{ss'}-n_xa_x\hat{x})} {\rm e}^{{\rm i}(k_z+g_z)\tilde{z}_{ss'}}, \label{Eq_1dlatsum}\\
&\tilde{\bm \rho}_{ss'}=\tilde{\bm \rho}_{s}-\tilde{\bm \rho}_{s'}, \quad \tilde{z}_{ss'}=\tilde{z}_{s}-\tilde{z}_{s'}, \quad \tilde{\bm x}_s=(\tilde{\rho}_s,\tilde{z}_s).
\end{align}
\end{widetext}
In Eq. (\ref{Eq_1dlatsum}), the prime in the real lattice sum over $n_x$ means that for $s=s'$,  $n_x=0$ is excluded, but for $s\ne s'$, $n_x=0$ is included.  This lattice sum can be efficiently calculated with the Ewald method \cite{Ohtaka:U:A::57:p2550-2568:1998,moroz2001exponentially}.

The $S$-matrix describes the light scattering by the monolayer. 
The scattering by an $N$-layer-thick slab can be obtained by a layer-by-layer construction \cite{Pendry-LEED-book}.

As an example, let us consider the 2D photonic band structure for the monolayer of the DBR pillars. The 2D band structure again consists of the true-guided modes outside the light cone $\omega<c\sqrt{k_x^2+k_z^2}$ and the quasi-guided modes inside the light cone $\omega >c\sqrt{k_x^2+k_z^2}$. The latter modes have finite life times by the mixing with th radiation continuum. 

The dispersion relation of the guided modes are obtained by  
\begin{align}
{\rm det}[1-t(k_z)G(k_x,k_z)]=0. 
\end{align}
On the other hand, the dispersion relation of the quasi-guided modes is given by the peak of the optical DOS \cite{Ohtaka:I:Y::70:p035109:2004}
\begin{align}
&\Delta\rho(\omega,k_x,k_z)=\frac{1}{2\pi}\frac{\partial}{\partial \omega} {\rm arg}({\rm det}[U]), \\
&U=\left( \begin{array}{cc}
\tilde{S}^{++} & \tilde{S}^{+-} \\
\tilde{S}^{-+} & \tilde{S}^{--} 
\end{array}\right),\\
&[\tilde{S}^{\zeta \zeta'}]_{{\bm g}^o{\bm g}'^o}^{\sigma\sigma'}=\sqrt{\gamma_{{\bm g}^o}}{\bm d}_{{\bm g}^o}^{\sigma\zeta} \tensor{S}_{{\bm g}_o{\bm g}'^o}^{\zeta\zeta'} {\bm d}_{{\bm g}'^o}^{\sigma'\zeta'} \frac{1}{\sqrt{\gamma_{{\bm g}'^o}}} \nonumber\\
&\hskip100pt  (\sigma,\sigma'=P,S; \quad \zeta,\zeta'=\pm),\\
&{\bm d}_{\bm g}^{P\pm} = \left(\begin{array}{l}
\pm \frac{\gamma_{\bm g}}{q}\frac{k_x+g_x}{\sqrt{(k_x+g_x)^2+(k_z+g_z)^2}}\\
-\frac{\sqrt{(k_x+g_x)^2+(k_z+g_z)^2}}{q}\\
\pm \frac{\gamma_{\bm g}}{q}\frac{k_z+g_z}{\sqrt{(k_x+g_x)^2+(k_z+g_z)^2}}
\end{array}\right),\\
&{\bm d}_{\bm g}^{S\pm} =  \left(\begin{array}{l}
\frac{k_z+g_z}{\sqrt{(k_x+g_x)^2+(k_z+g_z)^2}}\\
0\\
-\frac{k_x+g_x}{\sqrt{(k_x+g_x)^2+(k_z+g_z)^2}}
\end{array}\right),
\end{align}
where ${\bm g}^o$ represents the open diffraction channels. The matrix $U$ is shown to be unitary provided that the DBR pillars are lossless.

\begin{figure*}
\begin{center}
\centerline{
\includegraphics[width=0.45\textwidth]{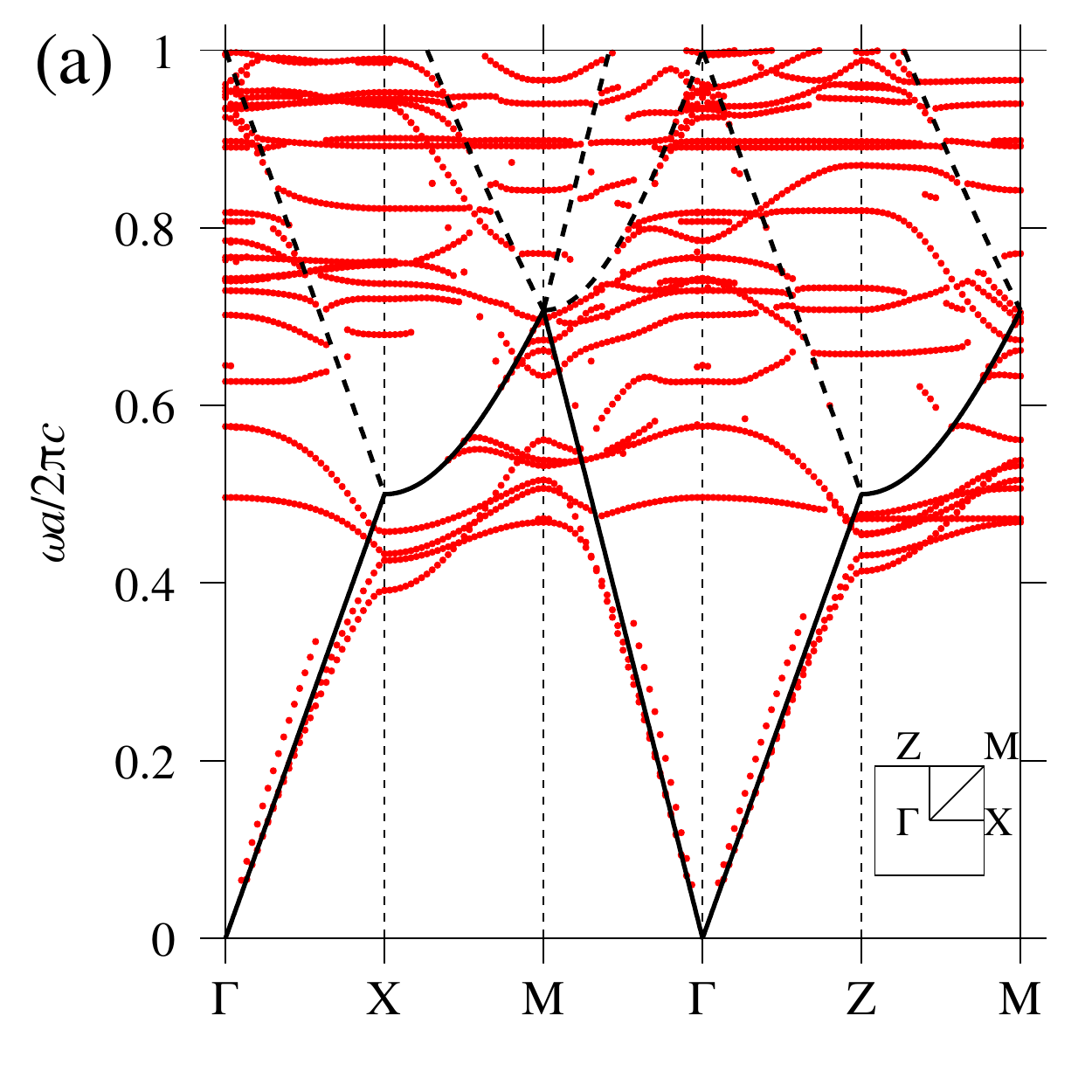}
\includegraphics[width=0.45\textwidth]{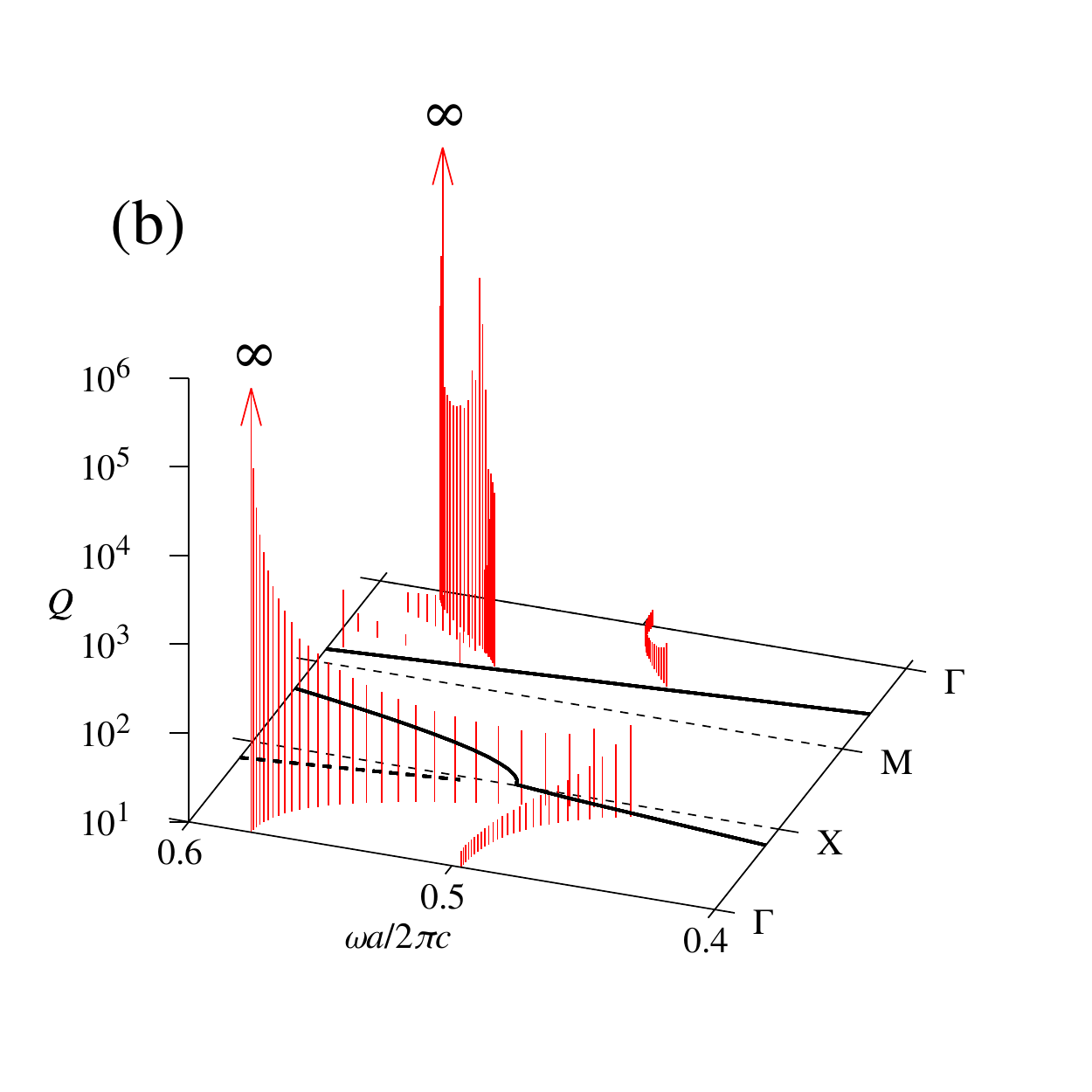}}
\end{center}
\caption{\label{Fig_band2d} (a) 2D photonic band structure of the simple monolayer of identical DBR pillars (only one pillar per unit cell). The Brillouin zone is shown in the inset. (b) Quality factor spectrum of the quasi-guided modes in a narrow frequency region (just to avoid complexity).  
The DBR parameters are the same as in Fig. \ref{Fig_band1d}. The lattice constant of the monolayer is equal to that of the DBR pillar [$a=a_x=a_z(=d)$]. The solid line represents the light line $\omega=c\sqrt{k_x^2+k_z^2}$. The dashed line represents the thresholds of the Bragg diffraction.     
}
\end{figure*}

A typical 2D band structure of the simple monolayer is shown in Fig. \ref{Fig_band2d}. 
We can see clearly the band structure formation inside and outside the light cone.  
The $Q$ values for the quasi-guided modes are also plotted in a narrow frequency window. Again, infinite $Q$ is obtained at the $\Gamma$ point. This is also the bound states in continuum due to the symmetry mismatch between the quasi-guided eigenmodes at $\Gamma$ and radiation continuum modes  there 
 \cite{Ochiai:S::63:p125107:2001}. In this system, we have the $C_{2v}$ symmetry in the $xz$ plane. Therefore, the eigenmodes are classified according to the parities $\sigma_x$ and $\sigma_z$.  The radiation continuum modes have either 
$(\sigma_x,\sigma_z)=(1,-1)$ or $(\sigma_x,\sigma_z)=(-1,1)$ below the Bragg-diffraction threshold $\omega a/2\pi c=1$.  The eigenmodes of $(\sigma_x,\sigma_z)=(1,1)$ and $(\sigma_x,\sigma_z)=(-1,-1)$ do not couple to the radiation continuum, so that they have infinite $Q$.

\subsection{2D periodic arrays}

In a 2D periodic array of DBR pillars, the so-called bulk-KKR method \cite{korringa1947ceb,kohn1954sse,leung1993msc} can be employed. The method allows us to study the photonic band structure in the bulk 3D PhC composed of DBR pillars. 

Suppose that the 2D lattice consists of a periodic array of aligned DBR pillars. Their central positions are denoted by 
\begin{align}
{\bm x}_\mu =\tilde{\bm x}_s + {\bm R}, \quad {\bm R}=n_1 {\bm e}_1+n_2{\bm e}_2 \quad (n_1,n_2\in{\bm Z}), 
\end{align}
where ${\bm e}_1$ and ${\bm e}_2$ are the elementary lattice vector in the $xy$ plane. In this case, the Bloch theorem is applied, so that the vector-cylindrical-wave-expansion coefficients of the induced radiation field satisfies 
\begin{align}
\psi_{\mu lg_z}^{\beta{\rm ind}}={\rm e}^{{\rm i}{\bm k}_\|\cdot{\bm R}} \chi_{slg_z}^{\beta{\rm ind}}.
\end{align}
Here, ${\bm k}_\|=(k_x,k_y)$ is the 2D Bloch momentum.

The secular equation for the bulk eigenmodes is given by 
\begin{align}
&\chi_{slg_z}^{\beta{\rm ind}}- \sum_{s'l'g_z'\beta'}[t_{sl}(k_z)]_{g_zg_z'}^{\beta\beta'}[G({\bm k})]_{ss';ll';g_z'}\chi_{s'l'g_z'}^{\beta'{\rm ind}}=0, \\
&[G({\bm k})]_{ss';ll';g_z}=\sum_{{\bm R}}{}' {\rm e}^{{\rm i}{\bm k}_\|\cdot{\bm R}}H_{l'-l}(\lambda_{g_z}|\tilde{\bm \rho}_{ss'}-{\bm R}|) \nonumber \\
&\hskip80pt \times 
{\rm e}^{{\rm i}(l'-l)\phi(\tilde{\bm \rho}_{ss'}-{\bm R})} {\rm e}^{{\rm i}(k_z+g_z)\tilde{z}_{ss'}}. \label{Eq_2dlatsum}
\end{align}
Here, the prime in the real lattice sum in Eq. (\ref{Eq_2dlatsum}) means for $s=s'$, ${\bm R}=0$ is excluded, whereas for $s\ne s'$, ${\bm R}=0$ is included.  This lattice sum can be efficiently evaluated by the Ewald method \cite{leung1993msc,Ohtaka:U:A::57:p2550-2568:1998}.

\begin{figure}
\begin{center}
\includegraphics[width=0.45\textwidth]{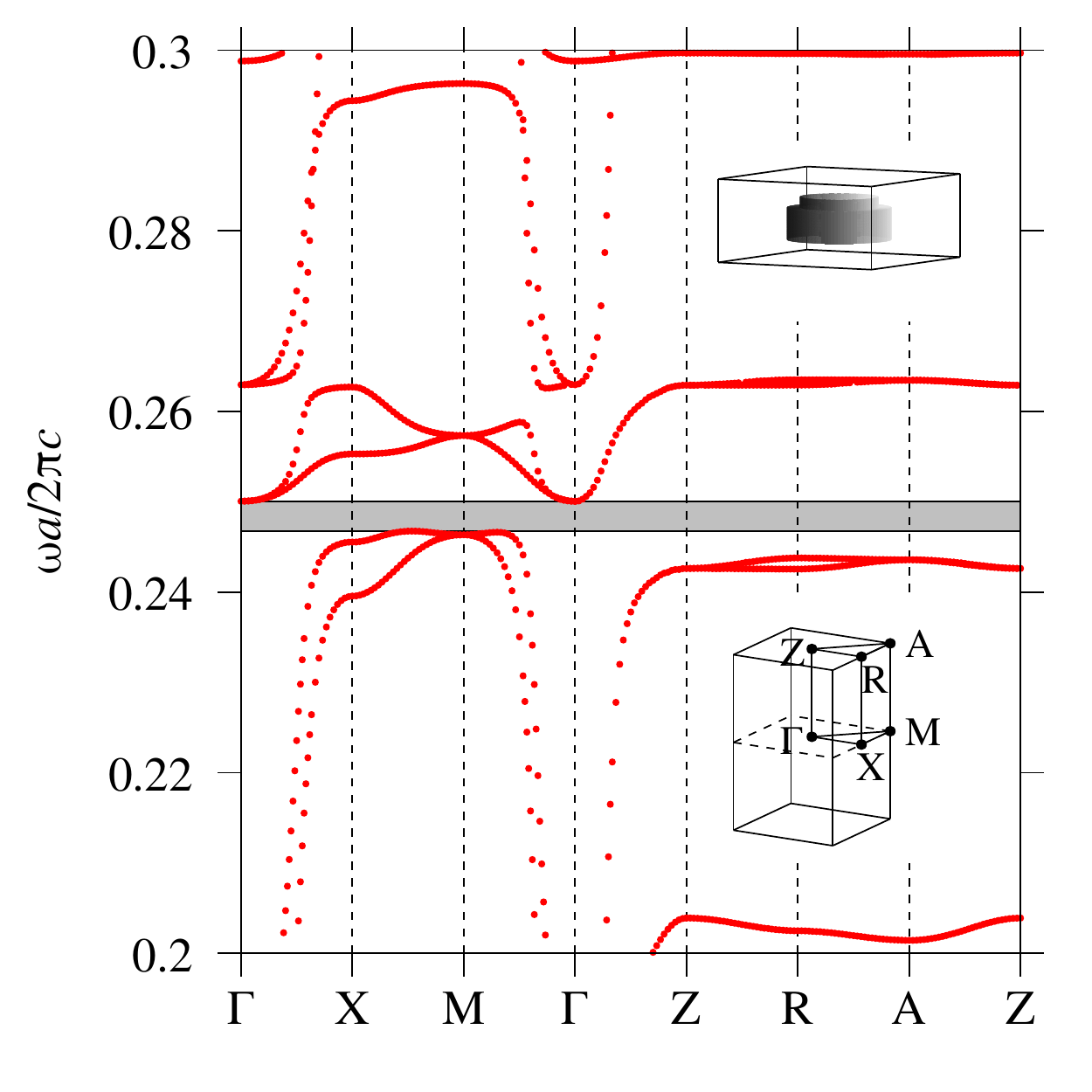}
\end{center}
\caption{\label{Fig_band3d} 3D photonic band structure of the square lattice of identical core-shell DBR pillars. A unit cell is shown in the upper inset. The Brillouin zone is shown in the lower inset. The gray stripe is the photonic band gap. 
The DBR parameters are as follows. The inner-core DBR has $\epsilon_a=100$, $\epsilon_b=1$, $d_a=0.259a$, $d_b=0.241a$, and $r_0=0.221a$. 
The outer-shell DBR has $\epsilon_a=100$, $\epsilon_b=1$, $d_a=0.194a$, $d_b=0.306a$, and $r_0=0.295a$. Here, $a$ is the lattice constant of the square lattice.}
\end{figure}

Here, we re-calculate the 3D photonic band structures of the systems studied in Ref. \onlinecite{PhysRevA.96.043842} (a 3D PhC composed of the square-lattice of core-shell DBR pillars), with the present formalism.  
The results are shown in Fig. \ref{Fig_band3d}. 
The results  reproduce the previously obtained results with the rigorous coupled-wave analysis (RCWA).  The computational speed here is generally faster than in the RCWA.   
In the gap around $\omega a/2\pi c=0.25$, gapless surface states emerge, as shown in the next section.

\section{Gapless surface states in certain 3D structures}

As a nontrivial application of the present formalism, we consider a novel formulation of gapless surface states in a 3D PhC composed of core-shell DBR pillars.

In an author's previous work, he presented a recipe to create gapless surface states in a certain class of  3D PhCs \cite{PhysRevA.96.043842}. It typically includes a 2D periodic array of core-shell DBR pillars.  There, we first considered a 3D tetragonal PhC composed of the square lattice of simple DBR pillars.  Then, we tuned system parameters (such as radius or height or dielectric constant of the finite-height pillars), in such a way that an accidental degeneracy takes place at  certain ${\bm k}$ points of high symmetry in the 3D Brillouin zone. After that, we introduced a core-shell structure that breaks a parity symmetry in the pillar axis, reducing the point group symmetry of the PhC from $D_{4h}$ to $C_{4v}$. This perturbation opens a gap and the gap was shown to support gapless surface states in a surface normal to the pillar axis. These results are well explained analytically by an effective hamiltonian and numerically by the RCWA.

However, the effective hamiltonian is highly anisotropic.  In one orientation of surfaces (normal to the pillar axis), we can easily show the effective hamiltonian has the solution of the gapless surface states. However. it is not the case for the other orientations. The RCWA also has the same tendency.   
Thus, it is not clear how the surface states in the other surface orientations look like.

\begin{figure}
\begin{center}
\includegraphics[width=0.45\textwidth]{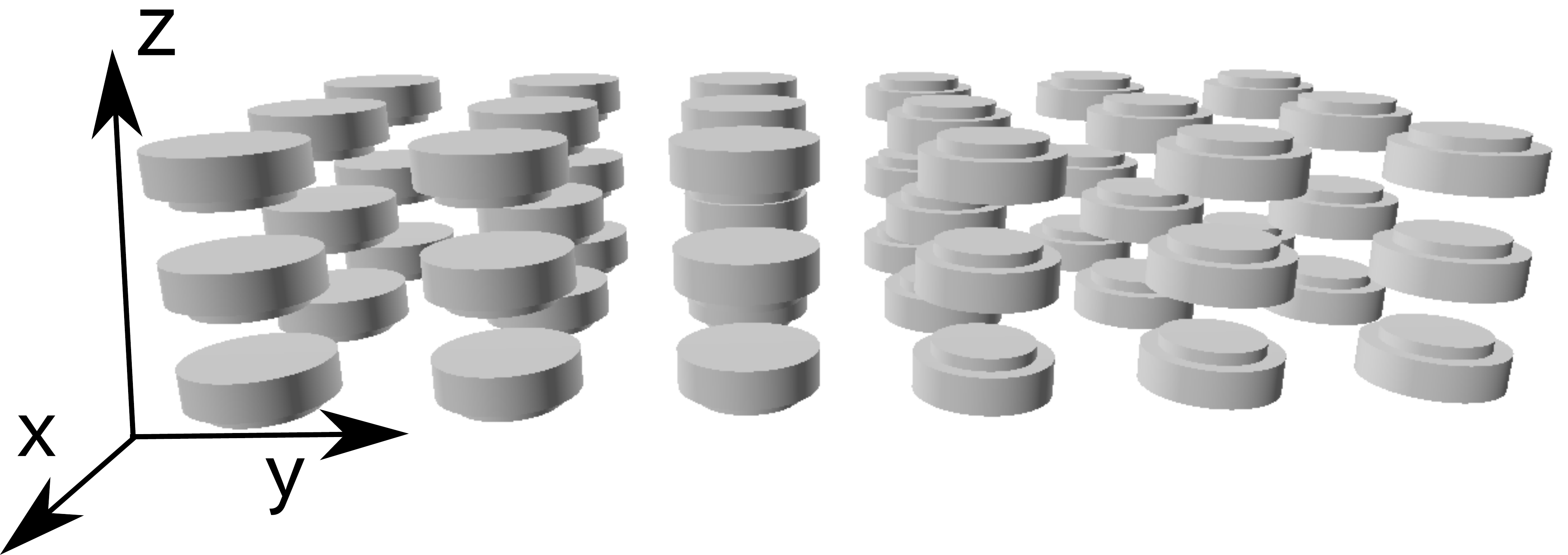}
\end{center}
\caption{\label{Fig_dw} Domain wall formed by two identical 3d tetragonal PhCs made of core-shell DBR pillars. In one domain, the $z$ axis in inverted from that in the other domain. The domain wall is normal to the $y$ axis. }
\end{figure}

To answer this question, we study the surface states in the tetragonal PhC of the core-shell DBR pillars in the present formalism.   
A schematic illustration of the system under study is shown in Fig. \ref{Fig_dw}. 
The surface (to be strict, domain-wall) state dispersion relation is obtained by solving the following secular equation: 
\begin{align}
{\rm det}(1-S_{\rm L}^{+-}S_{\rm R}^{-+})=0,  \label{Eq_sec_dw}
\end{align}
where $S_{\rm L(R)}^{\pm\pm}$ is the $S$-matrix of the left (right) domain. 
Ideally, we need the $S$-matrix of semi-infinite systems to screen the radiation field by the bulk band gap in the left and right domains, but in practice, $S$-matrices of finite-thick slabs are enough.

\begin{figure*}
\begin{center}
\centerline{
\includegraphics[width=0.45\textwidth]{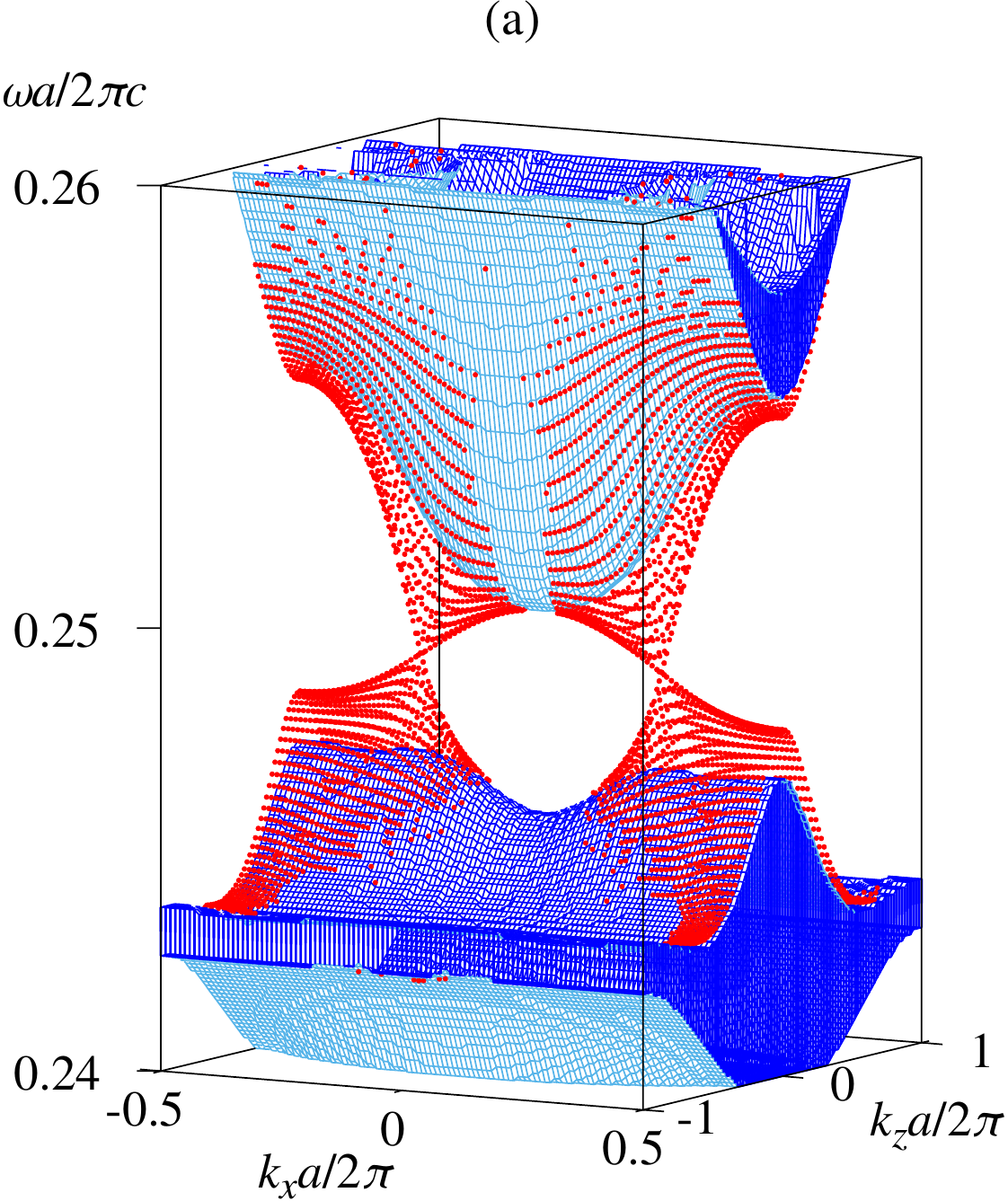}
\includegraphics[width=0.45\textwidth]{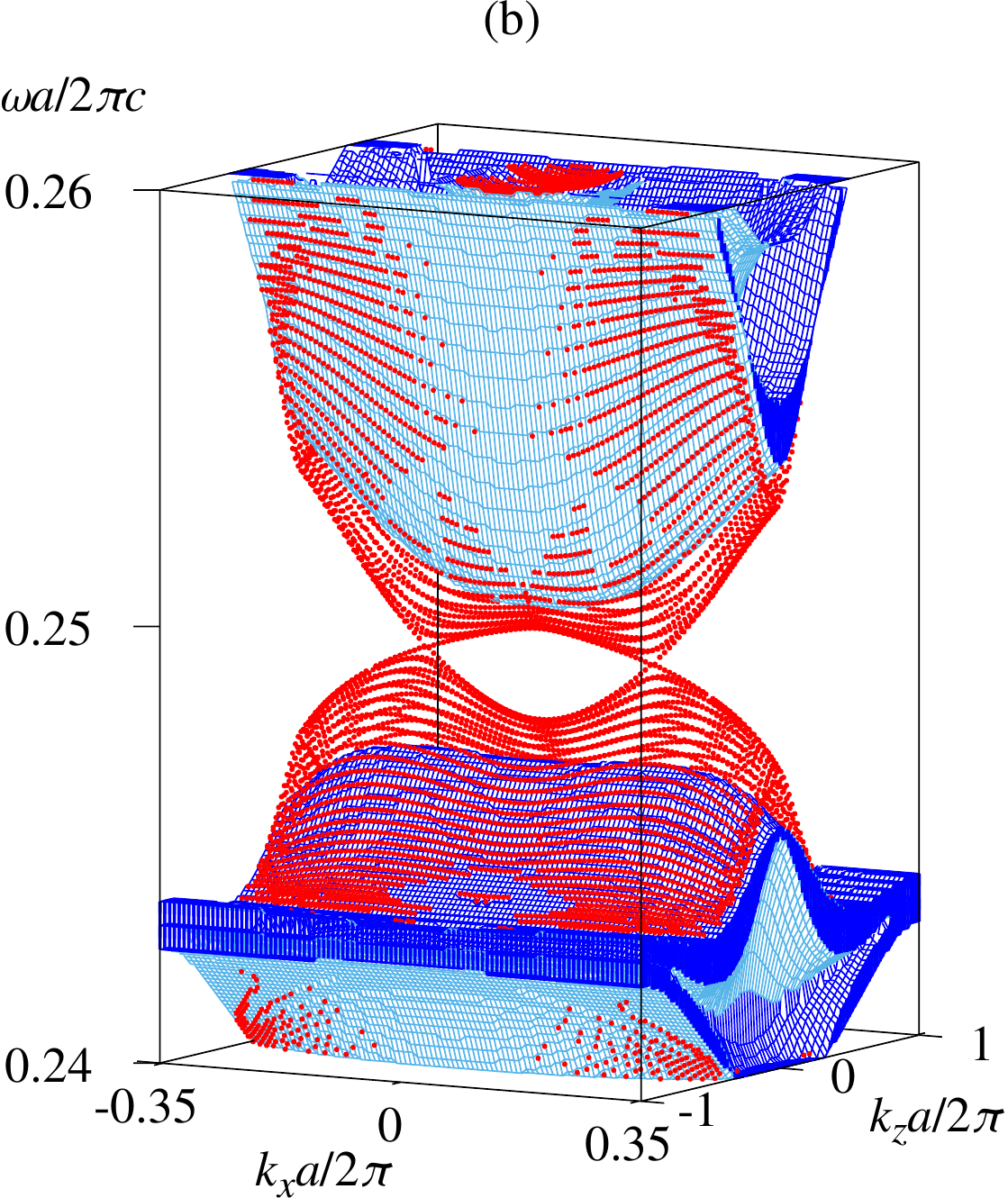}
}
\end{center}
\caption{\label{Fig_dirac} Dispersion relation of the domain-wall states in the 3D PhC composed of identical core-shell DBR pillars.  (a) Domain wall is normal to the $\Gamma X$ direction (see the lower inset of Fig. \ref{Fig_band3d}). (b) Domain wall is normal to the $\Gamma M$ direction. 
The DBR parameters are the same as in Fig. \ref{Fig_band3d}. The shaded region is the projection of the bulk band structure onto the surface Brillouin zone. 
}
\end{figure*}

We consider two surface orientations. One is normal to $\Gamma X$, and the other is normal to  $\Gamma M$. The results are shown in Fig. \ref{Fig_dirac}.
As seen clearly, we commonly have gapless surface states with two anisotropic Dirac cones. 
The Dirac points are located on $k_z=0$. 
The surface state dispersion merge with the band edge of the bulk band structure.

We also note the Dirac-cone dispersion does not change even after shifting the relative position between the two domains.  As far as we checked, the Dirac point just moves a bit depending on the relative position. The gaplessness of the surface states is unchanged. In this sense, the gapless surface states are robust against changing domain-wall profiles, like as in the gapless domain-wall fermion in the 2D Dirac hamiltonian \cite{PhysRevD.13.3398}.

To convince  the Dirac-cone formation, we consider the Berry phase of the surface states around the Dirac points. The Berry phase $\gamma_{\rm B}$ is defined as 
\begin{align}
&\gamma_{\rm B}=-{\rm i}\oint {\rm d}{\bm k}\cdot \left( (C_{\bm k}^+)^\dagger \frac{\partial}{\partial {\bm k}} C_{\bm k}^+\right), \label{Eq_Berry}\\
&C_{\bm k}^+ = {\rm Ker}(1-S_{\rm L}^{+-}S_{\rm R}^{-+}),\quad  (C_{\bm k}^+)^\dagger C_{\bm k}^+=1, 
\end{align}
where the integration path is a closed loop in the momentum space of the surface Brillouin zone.  
Here, the electric field of the domain-wall state is given by 
\begin{align}
&{\bm E}({\bm x})=\sum_{\bm g}\left[(c_{\bm g}^{P+}{\bm d}_{\bm g}^{P+}+c_{\bm g}^{S+}{\bm d}_{\bm g}^{S+}){\rm e}^{{\rm i}{\bm K}_{\bm g}^+\cdot{\bm x}} \right. \nonumber \\ 
&\hskip50pt \left.+ (c_{\bm g}^{P-}{\bm d}_{\bm g}^{P-}+c_{\bm g}^{S-}{\bm d}_{\bm g}^{S-}){\rm e}^{{\rm i}{\bm K}_{\bm g}^-\cdot{\bm x}} \right],\\
&C_{\bm k}^- = S_{\rm R}^{-+}C_{\bm k}^+,\\
&C_{\bm k}^\pm\equiv (c_{{\bm g}_1}^{P\pm},\dots,c_{{\bm g}_N}^{P\pm},c_{{\bm g}_1}^{S\pm},\dots,c_{{\bm g}_N}^{S\pm})^t,  
\end{align}
inside the void space between the two domains.   
In the actual calculation of $\gamma_{\rm B}$, we employ a gauge-invariant discretized version of Eq. (\ref{Eq_Berry}) \cite{fukui2005cnd}.

We found that the Berry phase converges to $\gamma_{\rm B}=\pm \pi$ if the loop contains the Dirac point inside and shrinks to it. 
Off the Dirac point,  the Berry phase converges to  $\gamma_{\rm B}=0$. It is well known the Berry phase around the Dirac point is $\pm \pi$ \cite{ando1998berry}.  Therefore, our system certainly has the Dirac points.

In our system, we have the time-reversal symmetry and at most parity symmetry $k_x\to -k_x$ depending on the relative position between the two domains. Thus, the Berry curvature $\Omega({\bm k})$ of the domain-wall state satisfies the following symmetry relations:  
\begin{align}
&\Omega(-k_x,-k_z) = -\Omega(k_x,k_z), \\ 
&\Omega(-k_x,k_z) = -\Omega(k_x,k_z),\\
&\Omega({\bm k})=-{\rm i}\left[\frac{\partial}{\partial {\bm k}}\times   \left( (C_{\bm k}^+)^\dagger \frac{\partial}{\partial {\bm k}} C_{\bm k}^+\right)\right]_y.
\end{align} 
These equations imply that the Berry curvature is generally nonzero, in contrast to  the systems with the both time-reversal symmetry and space-inversion symmetry, where the Berry curvature vanishes.  
However, if the loop shrinks to a point, then the Berry phase vanishes. 
An exception is that the point has some singularity. This is the case in our system, namely, we have a singularity of the Dirac point in a sense that the two bands become degenerate there.

The Berry phase of $\pm \pi$ implies a spin-momentum locking of the domain-wall state.  
Suppose that a certain reciprocal-lattice component ${\bm g}_{\rm d}$ is dominating in the domain-wall states.  Then, we can fairly neglect the other reciprocal-lattice components. Let us fix the gauge such that the $S$ polarization component is real and positive. Here, we refer the gauge transformation to as $C_{\bm k}^+\to C_{\bm k}^+\exp({\rm i}\varphi_{\bm k})$.  Then, we have  
\begin{align}
c_{{\bm g}_{\rm d}}^{P+}\simeq {\rm e}^{{\rm i}\alpha}\cos\phi, \quad c_{{\bm g}_{\rm d}}^{S+}\simeq \sin\phi.  
\end{align}  
The Berry curvature is now written as 
\begin{align}
\gamma_B\simeq\oint {\rm d}\alpha \cos^2\phi. 
\end{align}
Along the loop on which $P$- and $S$-polarized components contribute equally, we have $\cos^2\phi=1/2$, so that $\alpha$ winds $2\pi$. Namely, the $P$-polarized component relative to the $S$-polarized one has the winding feature in its argument. This phenomenon is a kind of the photospin-momentum locking, found in a 3D photonic topological insulator \cite{slobozhanyuk2016three}.

\section{Summary}

In summary, we have presented a hybrid method of the plane-wave and cylindrical-wave expansions for isolated DBR pillars.  After presenting fundamental properties of isolated DBR pillars, such as 1D photonic band structures, we have shown that the bound states in continuum emerge in the DBR pillars owing to a symmetry mismatch. 
 The method was then applied to periodic arrays of DBR pillars which form  2D or 3D PhCs. In particular, the $S$-matrices of 2D PhCs have been derived explicitly, and  a typical 2D band structure with Quality factors has been shown. Again, the bound states in continuum at the $\Gamma$ point are found.    
As a nontrivial application of the present formalism, we have presented a novel formation of gapless surface (domain-wall) states with two Dirac cones in a certain 3D PhC composed of core-shell DBR pillars.  
The presence of the Dirac cones are verified by calculating Berry curvature and  a possible photospin-momentum locking scenario has been derived.

\begin{acknowledgments}
This work was supported by JSPS KAKENHI Grant No. 17K05507. 
\end{acknowledgments}


%

\end{document}